\documentclass[aps,prb,twocolumn,showpacs,floatfix,longbibliography,superscriptaddress,10pt]{revtex4-1}
\usepackage{graphicx,amsfonts,amssymb,amsmath}
\usepackage{textcomp}
\usepackage{esint}
\usepackage[colorlinks,linktocpage,bookmarks=false,citecolor=blue,linkcolor=red,urlcolor=blue]{hyperref}
\usepackage[utf8]{inputenc}
\usepackage[T1]{fontenc}
\usepackage[dvipsnames]{xcolor}
\usepackage[english]{babel}
\usepackage{afterpage}
\usepackage{booktabs}

\vspace{0.3cm}
\allowdisplaybreaks

\def\Jmua{J^a_\mu} 
 
\def\jmua{j^a_\mu}  
 
\def\jnub{j^b_\nu} 
\def\Amu{A_\mu} 
\def\Anu{A_\nu} 
\def\Amua{A^a_\mu}

\def\Kmunuab{K^{ab}_{\mu\nu}}

\def\Pimunuab{\Pi^{ab}_{\mu\nu}}

\def\sigmunuab{\sig^{ab}_{\mu\nu}}

\def\Gamzdmunuab{\Gamma^{(0,2)ab}_{\mu\nu}}

\def\Gamuuimua{\Gamma^{(1,1)a}_{i\mu}}
\def\Gamuujmua{\Gamma^{(1,1)a}_{j\mu}}

\def\Gamuujmua{\Gamma^{(1,1)a}_{j\mu}}
\def\Gamuujnub{\Gamma^{(1,1)b}_{j\nu}}

\def\Cdis{C_{\rm dis}} 
\def\Lord{L_{\rm ord}}

\def\perm{\mbox{perm}} 
\def\sigA{\sig_{\rm A}}
\def\sigB{\sig_{\rm B}}
\def\sigAk{\sig_{{\rm A},k}}
\def\sigBk{\sig_{{\rm B},k}}
\def\sigQ{\sig^{}_Q}
\newcommand{\pv}{\mathbf{p}}
\newcommand{\qv}{\mathbf{q}}
\newcommand{\xiv}{{\boldsymbol \xi}}
\newcommand{\GL}{G_\mathrm{L}}
\newcommand{\GT}{G_\mathrm{T}}
\newcommand{\GLTk}{G_{\mathrm{L},\mathrm{T},k}}
\newcommand{\nnl}{\nonumber \\}
\newcommand{\rhos}{\rho_{s}}

\newcommand{\chiLT}{\chi_{\mathrm{L},\mathrm{T}}}
\newcommand{\chiL}{\chi_{\mathrm{L}}}
\newcommand{\chiT}{\chi_{\mathrm{T}}}
\newcommand{\chis}{\chi_{s}}

\newcommand{\Oc}{\mathcal{O}}

\newlength{\textlarg}

\def\beq{\begin{equation}}
\def\eeq{\end{equation}}
\def\bleq{\begin{eqnarray}}
\def\eleq{\end{eqnarray}} 
\def\bfig{\begin{figure}}
\def\efig{\end{figure}}
\def\bline{\begin{multline}}
\def\eline{\end{multline}}
\def\bremark{\begin{quotation} \noindent \small }
\def\eremark{\end{quotation}}
\def\llbrace{\left\lbrace}
\def\rrbrace{\right\rbrace}

\newcommand{\Tr}{{\rm Tr}} 
\newcommand{\tr}{{\rm tr}} 
 
\newcommand{\mean}[1]{\langle #1 \rangle}

\newcommand{\const}{{\rm const}}

\def\phibf{\boldsymbol{\phi}}
\def\varphibf{\boldsymbol{\varphi}}

\def\Lamb{\Lambda}
\def\sig{\sigma}
\def\Sig{\Sigma}

\def\half{\frac{1}{2}}

\def\quarter{\frac{1}{4}}

\def\p{{\bf p}} 
\def\q{{\bf q}}
\def\r{{\bf r}}

\def\x{{\bf x}}

\def\A{{\bf A}}

\def\D{{\bf D}}

\def\J{{\bf J}}

\def\w{\omega}
\def\wn{\omega_n}

\def\dmu{{\partial_\mu}}
\def\dnu{{\partial_\nu}}

\def\dk{\partial_k}

\def\inttau{\int_0^\beta d\tau}

\def\calA{{\cal A}}
\def\calB{{\cal B}}

\def\calD{{\cal D}}

\def\calO{{\cal O}}
  
\def\calR{{\cal R}}

\def\calZ{{\cal Z}}

\def\trho{{\tilde\rho}}

\begin{document}

\title{Nonperturbative renormalization-group approach preserving the momentum dependence of correlation functions}

\author{F. Rose}
\affiliation{Technische\,Universit{\"a}t\,M{\"u}nchen,\,Physik\,Department,\,James-Franck-Strasse,\,85748 Garching,\,Germany} 
\affiliation{Sorbonne Universit\'e, CNRS, Laboratoire de Physique Th\'eorique de la Mati\`ere Condens\'ee, LPTMC, F-75005 Paris, France}

\author{N. Dupuis}
\affiliation{Sorbonne Universit\'e, CNRS, Laboratoire de Physique Th\'eorique de la Mati\`ere Condens\'ee, LPTMC, F-75005 Paris, France}

\date{April 23, 2018} 

\begin{abstract}  
We present an approximation scheme of the nonperturbative renormalization group that preserves the momentum dependence of correlation functions. This approximation scheme can be seen as a simple improvement of the local potential approximation (LPA) where the derivative terms in the effective action are promoted to arbitrary momentum-dependent functions. As in the LPA the only field dependence comes from the effective potential, which allows us to solve the renormalization-group equations at a relatively modest numerical cost (as compared, e.g., to the Blaizot--Mend\'ez-Galain--Wschebor approximation scheme). As an application we consider the two-dimensional quantum O($N$) model at zero temperature. We discuss  not only the two-point correlation function but also higher-order correlation functions such as the scalar susceptibility (which allows for an investigation of the ``Higgs'' amplitude mode) and the conductivity. In particular we show how, using Pad\'e approximants to perform the analytic continuation $i\wn\to\w+i0^+$ of imaginary frequency correlation functions $\chi(i\wn)$ computed numerically from the renormalization-group equations, one can obtain spectral functions in the real-frequency domain. 
\end{abstract}

\maketitle

\tableofcontents

\section{Introduction}

The nonperturbative renormalization-group (NPRG) provides us with a general formalism to study classical and quantum many-body systems.\cite{Berges02,Delamotte12,Kopietz_book} It has been applied to a variety of physical systems ranging from particle-physics to statistical mechanics and condensed matter (see, e.g., Ref.~\onlinecite{Berges02}). 

The NPRG is based on an exact flow equation for the effective action (or Gibbs free energy), a functional of the order parameter.\cite{Wetterich93,Ellwanger93,Morris94} In general, this equation cannot be solved but offers the possibility of approximation schemes qualitatively different from perturbation theory, allowing, in particular, to tackle nonperturbative problems. So far two main approximations have been proposed. The first one relies on a derivative expansion (DE) of the effective action.\cite{Tetradis94,Morris99,Gersdorff01,Canet03a,Jakubczyk14} A nice feature of this approach is the possibility to implement the various symmetries of the problem rather easily. One of its main drawbacks is that it gives access to correlation functions only at vanishing momenta. It can also break down due to some vertices being singular in the infrared limit. In that case even the zero-momentum value of correlation functions is out of reach. The second one, the Blaizot--M\'endez-Galain--Wschebor (BMW) approximation scheme,\cite{Blaizot06,Benitez09,Benitez12} is based on a truncation of the infinite hierarchy of equations satisfied by correlation functions. Its main advantage over DE is to preserve the full momentum dependence of (low-order) correlation functions. Its limitations are twofold. First it leads to flow equations which, in some cases, can be solved only at a high numerical cost. Second, symmetries can be difficult to implement. 

In this paper we consider another approximation scheme, dubbed LPA$''$ for reasons that will become clear below (LPA stands for local potential approximation). The LPA$''$ was originally introduced in Ref.~\onlinecite{Hasselmann12} to compute the critical exponents and momentum-dependent correlation functions in the O($N$) model. By contrast with the DE, the LPA$''$ relies on an ansatz for the effective action parameterized by \emph{non-local} potentials, an idea that has been recently discussed both in the context of statistical physics\cite{Canet16a} and quantum field theory.\cite{Feldmann17} It was used in Ref.~\onlinecite{Rose17a} for the calculation of the conductivity of the two-dimensional quantum O($N$) model in order to circumvent the failure of BMW and DE approximation schemes.\footnote{Momentum- and frequency-dependent correlation functions have also been studied in the context of QCD: see, e.g., Refs.~\onlinecite{Kamikado14,Tripolt14,Tripolt14a,Wambach14,Pawlowski15,Cyrol18,Pawlowski17}}

Our aim is to benchmark the LPA$''$ considering as a test-bed the two-dimensional quantum O($N\geq 2$) model at zero temperature. In addition to the critical exponents of the quantum phase transition due to the spontaneous breaking of O($N$) symmetry, the excitation gap in the disordered phase and the stiffness in the ordered phase, we compute the momentum dependence of the two-point correlation function, the scalar O($N$) invariant susceptibility (which allows for an investigation of the ``Higgs'' amplitude mode\cite{Podolsky11}) and the conductivity. Since at zero temperature the two-dimensional quantum model is equivalent to the three-dimensional classical model, we shall in a first step consider the latter and compute the momentum dependence of the various correlation functions $\chi(\p)$ of interest. To obtain the retarded correlation functions and the spectral functions in the two-dimensional quantum model we then perform an analytic continuation $|\p|\to -i\w+0^+$ using Pad\'e approximants.\cite{Vidberg77} 

The outline of the paper is as follows. The general formalism is introduced in Sec.~\ref{sec_nprg}. After a presentation of the quantum O($N$) model (Sec.~\ref{subsec_qON}) and the NPRG approach to the computation of the two-point correlation function, scalar susceptibility and conductivity (Sec.~\ref{subsec_nprg}), we describe the LPA$''$ (Sec.~\ref{subsec_lpapp}). Results for universal quantities near the quantum critical point (QCP), critical exponents and universal scaling functions, are discussed in Sec.~\ref{sec_spectral}. Whenever possible comparison is made with DE and BMW results as well as Monte Carlo simulations or conformal bootstrap. Technical details can be found in Appendix~\ref{app_rgeq}.

\section{NPRG approach} 
\label{sec_nprg}
 
\subsection{Quantum O($N$) model} 
\label{subsec_qON}
 
The two-dimensional quantum O($N$) model is defined by the Euclidean action
\begin{equation}
S = \int_\x \biggl\lbrace \half \sum_{\mu=x,y,\tau}(\dmu\varphibf)^2  
+ \frac{r_0}{2} \varphibf^2 + \frac{u_0}{4!N} {(\varphibf^2)}^2  \biggr\rbrace ,
\label{action1} 
\end{equation}
where we use the notation $\x=(\r,\tau)$ and $\int_\x=\inttau \int d^2r$. $\varphibf(\x)$ is an $N$-component real field,  $\r$ a two-dimensional coordinate, $\tau\in [0,\beta]$ an imaginary time, and $\beta=1/T$ the inverse temperature (we set $\hbar=k_B=1$). $r_0$ and $u_0$ are temperature-independent coupling constants and the (bare) velocity of the $\varphibf$ field has been set to unity. The model is regularized by an ultraviolet cutoff $\Lamb$. Assuming $u_0$ fixed, there is a quantum phase transition between a disordered phase ($r_0>r_{0c}$) and an ordered phase ($r_0<r_{0c}$) where the O($N$) symmetry is spontaneously broken. The QCP at $r_0=r_{0c}$ is in the universality class of the three-dimensional classical O($N$) model and the phase transition is governed by the three-dimensional Wilson--Fisher fixed point. 

At zero-temperature the two-dimensional quantum model is equivalent to the three-dimensional classical model. We thus identify $\tau$ with a third spatial dimension so that $\x=(\r,\tau)\equiv (x,y,z)$. A correlation function $\chi(p_x,p_y,p_z)$ computed in the classical model then corresponds to the correlation function $\chi(p_x,p_y,i\wn)$ of the quantum model, with $\wn\equiv p_z$ a bosonic Matsubara frequency,\footnote{At zero temperature, the bosonic Matsubara frequency $\wn=2n\pi T$ ($n$ integer) becomes a continuous variable.} and yields the retarded dynamical correlation function $\chi^R(p_x,p_y,\w)$ after analytic continuation $i\wn\to\w+i0^+$. Having in mind the two-dimensional quantum O($N$) model, we shall refer to the critical point of the three-dimensional classical model as the QCP.

\subsubsection{Scalar susceptibility} 
\label{sec_Sh} 

To compute the scalar, O($N$) invariant, susceptibility
\beq 
\chis(\x,\x') = \mean{\varphibf(\x)^2 \varphibf(\x')^2} -\mean{\varphibf(\x)^2} \mean{\varphibf(\x')^2} ,
\label{chis_def} 
\eeq
we introduce an external source term $h$ which couples to $\varphibf^2$,
\beq
S[\varphibf,h] = S[\varphibf] - \int_\x h \varphibf^2 . 
\label{Sh}
\eeq 
The scalar susceptibility can then be computed as the functional derivative 
\beq 
\chis(\x,\x') = \frac{\delta^2 \ln \calZ[h]}{\delta h(\x) \delta h(\x')} \biggl|_{h=0} 
\label{chis_def1} 
\eeq 
of the partition function $\calZ[h]=\int \calD[\varphibf] \exp(-S[\varphibf,h])$ in the presence of the source $h$.

\subsubsection{Conductivity}
\label{sec_SA} 

The O($N$) symmetry of the action~(\ref{action1}) implies the conservation of the total angular momentum and the existence of a conserved current. To compute the associated conductivity, we include in the model an external non-Abelian gauge field $A_\mu=A_\mu^a T^a$ (with an implicit sum over repeated discrete indices), where $\{T^a\}$ denotes a set of SO($N$) generators (made of $N(N-1)/2$ linearly independent skew-symmetric matrices). This amounts to replacing the derivative $\dmu$ in Eq.~(\ref{action1}) by the covariant derivative $D_\mu=\dmu-q A_\mu$ (we set the charge $q$ equal to unity in the following and restore it, as well as $\hbar$, whenever necessary),
\beq
S[\varphibf,\A] = \int_\x \biggl\lbrace \half \sum_{\mu=x,y,z}(D_\mu\varphibf)^2  
+ \frac{r_0}{2} \varphibf^2 + \frac{u_0}{4!N} {(\varphibf^2)}^2  \biggr\rbrace .
\label{SA} 
\eeq
This makes the action invariant in the local gauge transformation $\varphibf'=O\varphibf$ and $\Amu'= O \Amu O^T +(\dmu O)O^T$ where $O$ is a space-dependent SO($N$) rotation. 
The current density $\Jmua(\x) = - \delta S/\delta \Amua(\x)$ is then expressed as~\cite{Rose17} 
\beq
\Jmua =  \jmua -\Amu\varphibf \cdot T^a \varphibf ,  \quad
\jmua = \dmu \varphibf \cdot T^a \varphibf ,
\eeq
where $\jmua$ denotes the ``paramagnetic'' part. 

In the two-dimensional quantum model, the real-frequency conductivity is defined by\cite{Rose17} 
\begin{equation}
\sigmunuab(\w) = \sigmunuab(i\wn\to \w+i0^+) = \frac{1}{i(\w+i0^+)} \Kmunuab{}^R(\w) ,
\label{sigR}
\end{equation}
where $\Kmunuab{}^R(\w)=\Kmunuab(i\wn\to\w+i0^+)$ denotes the retarded part of $\Kmunuab(i\wn)\equiv\Kmunuab(p_x=0,p_y=0,p_z=\wn)$. $\Kmunuab(\p)$ is the correlation function of the three-dimensional classical O($N$) model defined by 
\beq
\Kmunuab(\x-\x') = \Pimunuab(\x-\x') - \delta_{\mu\nu}\delta(\x-\x')\mean{T^a\varphibf\cdot T^b\varphibf},
\label{Kdef2}
\eeq 
with
\begin{equation}
\Pimunuab(\x-\x') = \mean{\jmua(\x) \jnub(\x')} 
\end{equation} 
the paramagnetic current-current correlation function.

\subsection{NPRG formalism} 
\label{subsec_nprg} 

Let us briefly recall the main steps of the NPRG implementation (we refer to Ref.~\onlinecite{Rose17} for more detail).We add to the action a regulator term $\Delta S_k[\varphibf]$ which depends on a cutoff function $R_k(\q)$, where $k$ a momentum scale which varies from the microscopic scale $\Lamb$ down to 0.\cite{Berges02,Delamotte12,Kopietz_book} In practice we take the exponential cutoff function 
\beq
R_k(\q) = Z_k \q^2 r\left( \frac{\q^2}{k^2} \right), \quad r(y) = \frac{\alpha}{e^y-1} ,
\label{Rkdef} 
\eeq 
where $\alpha$ is a constant of order one and $Z_k$ a field-renormalization factor. (In the LPA$''$ discussed below, $Z_k\equiv Z_k(\p=0)$.)

The partition function $\calZ_k[\J]$, computed in the presence of an external source $\J$ linearly coupled to the field, is now $k$ dependent and so is the order parameter $\phibf_k[\x;\J] = \mean{\varphibf(\x)}$. The scale-dependent effective action $\Gamma_k[\phibf]$, defined as a (slightly modified) Legendre transform of $-\ln \calZ_k[\J]$, satisfies Wetterich's equation\cite{Wetterich93} 
\begin{equation}
\dk \Gamma_k[\phibf] = \half \Tr \Bigl\{ \dk R_k \bigl(\Gamma_k^{(2)}[\phibf] + R_k \bigr)^{-1} \Bigr\} ,
\label{Weteq}
\end{equation}
with initial condition $\Gamma_\Lambda[\phibf]=S[\phibf]$. At $k=0$  the regulator vanishes and $\Gamma_{k=0}[\phibf]=\Gamma[\phibf]$. Here $\Gamma_k^{(2)}[\phibf]$ denotes the second-order functional derivative with respect to $\phibf$ of $\Gamma_k[\phibf]$ and the trace runs over both space and internal O($N$) variables. 

All information about the thermodynamics of the system can be deduced from the effective potential 
$U_k(\rho) = V^{-1} \Gamma_k[\phibf]$ obtained from the effective action in a uniform field configuration ($V$ denotes the volume of the system).  
For symmetry reasons, $U_k$ is function of the O($N$) invariant $\rho=\phibf^2/2$. We denote by $\rho_{0,k}$ the value of $\rho$ at the minimum of the effective potential. Spontaneous symmetry breaking of the O($N$) symmetry is characterized by a nonvanishing expectation value of the field $\varphibf$, i.e., $\lim_{k\to 0}\rho_{0,k} = \rho_0 > 0$. 

On the other hand correlation functions can be related to the one-particle irreducible (1PI) vertices $\Gamma_k^{(n)}$ defined as the functional derivatives of $\Gamma_k$. In particular, the two-point correlation function (propagator) $G_k=(\Gamma_k^{(2)}+R_k)^{-1}$ is simply related to the two-point vertex $\Gamma_k^{(2)}$. The latter can be written as 
\beq
\Gamma^{(2)}_{k,ij}(\p,\phibf) = \delta_{ij} \Gamma_{A,k}(\p,\rho) + \phi_i \phi_j \Gamma_{B,k}(\p,\rho) ,
\label{Gamma2} 
\eeq 
where $\Gamma_{\rm A,k}$ and $\Gamma_{B,k}$ are functions of $\rho$ and $|\p|$. 
Important information can be obtained from the longitudinal and transverse susceptibilities, $\chi_\alpha(\p)\equiv G_{k=0,\alpha}(\p,\rho_0)$ ($\alpha={\rm L,T}$), where $G_{k,\alpha}^{-1}(\pv,\rho)=\Gamma^{(2)}_{\alpha,k}(\p,\rho)+R_k(\p)$ with
\begin{align}
\Gamma^{(2)}_{\mathrm{L},k}(\p,\rho)&=\Gamma_{A,k}(\p,\rho) + 2 \rho \Gamma_{B,k}(\p,\rho),\nnl
\Gamma^{(2)}_{\mathrm{T},k}(\p,\rho)&=\Gamma_{A,k}(\p,\rho).
\label{chi_def}
\end{align}
In the disordered phase ($\rho_0=0$), $\chi_{\rm L}(\p)=\chi_{\rm T}(\p)\equiv\chi_{\rm L,T}(\p)$ and the correlation length $\xi$ (i.e. the inverse of the excitation gap $\Delta$ of the quantum model\footnote{Lorentz invariance of the quantum model ensures that the velocity is not renormalized and equal to one in our units}) is finite. In the ordered phase, the stiffness $\rhos$ is defined by\cite{Chaikin_book,not1} 
\begin{equation}
\chi_{\rm T}(\p) =  \frac{2\rho_0}{\rhos \p^2} \quad \mbox{for} \quad \p \to 0 . 
\label{rhosdef}
\end{equation}
For two systems located symmetrically wrt the QCP (i.e. corresponding to the same value of $|r_0-r_{0c}|$), one in the ordered phase (with stiffness $\rhos$) and the other in the disordered phase (with correlation length $\xi=1/\Delta$), the ratio $\rhos/\Delta=\rhos\xi$ is a universal number which depends only on $N$. This allows us to use $\Delta$ as the characteristic energy scale in both the disordered and ordered phases (in the latter case, $\Delta$ is defined as the excitation gap at the point located symmetrically wrt the QCP).\cite{Podolsky12} $\Delta$ and $\rhos$ vanish as $|r_0-r_{0c}|^{\nu}$ as we approach the QCP. 

Although in principle the knowledge of the propagator $G_k$ and the four-point vertex $\Gamma_k^{(4)}$ is sufficient to obtain the scalar susceptibility $\chis$ and the conductivity $\sigma$, this approach is in practice difficult as it requires to know the momentum dependence of $\Gamma_k^{(4)}(\p_1,\p_2,\p_3,\p_4)$ for all momentum scales. It is much easier to compute $\chis$ and $\sigma$ directly from flow equations by introducing appropriate external sources as described in Secs.~\ref{sec_Sh} and \ref{sec_SA}.

\subsubsection{Scalar susceptibility}
\label{sec_nprg:subsubsec_chis} 

To compute the scalar susceptibility one considers the partition function $\calZ_k[\J,h]$ in the presence of both the linear source $\J$ and the bilinear source $h$.\cite{Rancon14,Rose15} The order parameter $\phi_k[\x;\J,h]$ is now a functional of both $\J$ and $h$. The scale-dependent effective action is defined as a Legendre transform wrt the source $\J$ (but not $h$) and satisfies the flow equation 
\begin{equation}
\dk \Gamma_k[\phibf,h] = \half \Tr \Bigl\{ \dk R_k \bigl(\Gamma_k^{(2,0)}[\phibf,h] + R_k \bigr)^{-1} \Bigr\} 
\label{Weteq_h}
\end{equation}
with the initial condition $\Gamma_\Lambda[\phibf,h]=S[\phibf]-\int_\x h\phibf^2$. $\Gamma_k^{(2,0)}[\phibf,h]$ denotes the second-order functional derivative of $\Gamma_k[\phibf,h]$ wrt $\phibf$. Using~(\ref{chis_def1}) one can relate the scalar susceptibility 
\begin{align}
\chis(\p) ={}& - \Gamma^{(0,2)}(\p,\bar\phibf) \nonumber \\ 
& + \Gamma_i^{(1,1)}(\p,\bar\phibf) \Gamma_{ij}^{(2,0)-1}(\p,\bar\phibf) \Gamma_j^{(1,1)}(\p,\bar\phibf) 
\label{chis_def2} 
\end{align}  
to the $k=0$ 1PI vertices $\Gamma^{(n,m)}$ defined as functional derivatives wrt to $\phi$ and $h$ (e.g. 
$\Gamma^{(1,1)}=\delta^2 \Gamma[\phi,h]/\delta\phi \delta h$) evaluated in a uniform field configuration.\cite{Rose15} In Eq.~(\ref{chis_def2}) $\bar\phibf$ denotes the (uniform) order parameter for $\J=h=0$ and we use the notation $\Gamma^{(n,m)}(\p)\equiv\Gamma^{(n,m)}(\p,-\p)$ for vertices with $n+m=2$. 

Using $\Gamma^{(1,1)}_i(\p,\phibf)=\phi_i f(\p,\rho)$ and $\Gamma^{(0,2)}(\p,\phibf)=\gamma(\p,\rho)$,  where $f$ and $\gamma$ are functions of $|\p|$ and $\rho$,\cite{Rose17} we obtain 
\beq
\chis(\p) = - \gamma(\p,\rho_0) + 2\rho_0 f(\p,\rho_0)^2 G_{\rm L}(\p,\rho_0) .
\eeq 
To determine the scalar susceptibility in the NPRG approach we must therefore consider the $k$-dependent vertices $\Gamma_k^{(0,2)}$ and $\Gamma_{k,i}^{(1,1)}$ or, equivalently, the $k$-dependent functions $f_k(\p,\rho)$ and $\gamma_k(\p,\rho)$, in addition to the effective potential $U_k(\rho)$ and the vertices $\Gamma_{A,k}(\p,\rho)$ and $\Gamma_{B,k}(\p,\rho)$ determining the propagator.

\subsubsection{Conductivity}
\label{sec_NPRG:subsubsec_cond} 

The conductivity can be calculated in a similar way.\cite{Rose17} However, to respect local gauge invariance, one must use the gauge-invariant regulator term $\Delta S_k[\varphibf,\A]$ obtained from $\Delta S_k[\varphibf]$ by replacing the derivative $\dmu$ by the covariant derivative $D_\mu$. The scale-dependent effective action 
is defined as a Legendre transform wrt the source $\J$ (but not $\A$) and satisfies the flow equation
\begin{equation}
\dk \Gamma_k[\phibf,\A] = \half \Tr \Bigl\{ \dk \calR_k[\A] \bigl(\Gamma_k^{(2,0)}[\phibf,\A] + \calR_k[\A] \bigr)^{-1} \Bigr\} ,
\label{Weteq_A}
\end{equation}
where $\Gamma_k^{(2,0)}[\phibf,\A]$ and $\calR_k[\A]$ denote the second-order functional derivative with respect to $\phibf$ of $\Gamma_k[\phibf,\A]$ and $\Delta S_k[\phibf,\A]$, respectively. 

One can relate the linear response
\begin{multline}
\Kmunuab(\p) = - \Gamzdmunuab(\p,\bar\phibf) \\ 
+ \Gamuuimua(-\p,\bar\phibf) \Gamma^{(2,0)-1}_{ij}(\p,\bar\phibf) \Gamuujnub(\p,\bar\phibf) 
\label{Kdef3} 
\end{multline} 
to the $k=0$ 1PI vertices $\Gamma^{(n,m)}$ defined as functional derivatives wrt $\phi$ and $A_\mu$ (e.g. $\Gamuujmua=\delta^2 \Gamma/\delta\phi_j\delta A_\mu^a$) 
computed in a uniform field $\phibf$ and for $\A=0$.\cite{Rose17} In Eq.~(\ref{Kdef3}) $\bar\phibf$ is the (uniform) order parameter in the absence of the gauge field. The O($N$) symmetry implies that 
\begin{equation}
\begin{split} 
\Gamuujmua(\p,\phibf) ={}& ip_\mu (T^a\phibf)_j \Psi_A, \\ 
\Gamzdmunuab(\p,\phibf) ={}& p_\mu p_\nu [ \delta_{ab} \Psi_B + (T^a\phibf)\cdot(T^b\phibf)\Psi_C ] \\ 
& + \delta_{\mu\nu} [ \delta_{ab} \bar\Psi_B + (T^a\phibf)\cdot(T^b\phibf)\bar\Psi_C ],
\end{split}
\label{vertA} 
\end{equation}
where $\Psi_A,\Psi_B,\Psi_C,\bar\Psi_B,\bar\Psi_C$ are functions of $\rho$ and $|\p|$. These functions are not independent but are related by Ward identities.\cite{Rose17} 

To obtain the frequency-dependent conductivity in the quantum model, one sets $\p=(0,0,\wn)$  and $\mu,\nu\in\{x,y\}$ so that $p_\mu=p_\nu=0$ and 
$\Kmunuab(i\wn) = -\Gamzdmunuab(i\wn,\bar\phibf)$ is fully determined by $\Psi_B(i\wn)$ and $\Psi_C(i\wn)$. In the disordered phase ($\rho_0=\bar\phibf^2/2=0$), the conductivity tensor $\sigma_{\mu\nu}^{ab}(i\wn)=\delta_{\mu\nu}\delta_{ab}\sigma(i\wn)$ is diagonal with
\beq
\sig(i\wn) = -\wn \Psi_B(i\wn,\rho_0) .  
\eeq 
In the ordered phase, when $N \geq 3$, the conductivity tensor is defined by two independent components\cite{Podolsky11} $\sigA(i\wn)$ and $\sigB(i\wn)$ such that
\begin{align}
\sigma_{\mu\nu}^{ab}(i\wn)={}&{}\delta_{\mu\nu}\bigg\{\dfrac{(T^a\phibf)\cdot(T^b\phibf)}{2\rho}[\sigA(i\wn)-\sigB(i\wn)] \nonumber\\
{}&{}+\delta_{ab}\sigB(i\wn)\bigg\}
\end{align}
with
\beq
\begin{split}
\sigA(i\wn) ={}& \frac{2\rho_0}{\wn} \Psi_A(i\wn,\rho_0) \\ 
& - \wn [ \Psi_B(i\wn,\rho_0) + 2\rho_0  \Psi_C(i\wn,\rho_0) ]   , \\ 
\sigB(i\wn) ={}& -\wn \Psi_B(i\wn,\rho_0) .
\end{split}
\label{sigABdef} 
\eeq
For $N=2$ there is only one SO$(N)$ generator and $\sigma$ reduces to $\sigA$.

\subsection{LPA$''$}
\label{subsec_lpapp}

The flow equations~(\ref{Weteq}), (\ref{Weteq_h}) and (\ref{Weteq_A}) cannot be solved exactly and one has to resort to approximations. In this section we discuss the LPA$''$, first for the calculation of the two-point correlation function ($\A=h=0$) and then for the scalar susceptibility and the conductivity. 

\subsubsection{Two-point correlation function} 

The LPA$''$ can be seen as an improvement of the LPA, where the ansatz for the effective action
\beq
\Gamma_k^{{\rm LPA}}[\phibf] = \int_{\x} \llbrace \half  \dmu\phibf\cdot\dmu\phibf + U_k(\rho) \rrbrace 
\label{gamlpa}
\eeq 
depends only on the effective potential $U_k(\rho)$. In the LPA$'$, the ansatz
\begin{align}
\Gamma_k^{{\rm LPA}'}[\phibf] ={}& \int_{\x} \biggl\lbrace \half (\dmu\phibf) \cdot Z_k (\dmu\phibf) \nonumber \\ 
& + \quarter (\partial_\mu \rho) Y_k (\partial_\mu \rho) + U_k(\rho) \biggr\rbrace 
\label{gamlpap}
\end{align} 
includes a field-renormalization factor $Z_k$ and (sometimes) a derivative quartic term $Y_k$. The standard improvement of the LPA$'$ is the derivative expansion to second order where $Z_k$ and $Y_k$ become functions of $\rho$.\cite{Berges02} Here we follow a different route and improve over the LPA$'$ by promoting $Z_k$ and $Y_k$ to functions of the derivative $-{\boldsymbol\partial}^2\equiv -\partial_\mu^2$, which yields 
\begin{align}
\Gamma_k^{{\rm LPA}''}[\phibf] ={}& 
 \int_\x \biggl\lbrace  \half (\dmu\phibf) \cdot Z_k(-{\boldsymbol\partial}^2) (\dmu\phibf) 
 \nonumber \\ & + \quarter (\partial_\mu \rho) Y_k(-{\boldsymbol\partial}^2) (\partial_\mu \rho) + U_k(\rho) \biggr\rbrace 
\label{lpapp} 
\end{align}
with initial conditions $Z_\Lamb(\p)=1$, $Y_\Lamb(\p)=0$ and $U_\Lamb(\rho)=r_0\rho+(u_0/6N)\rho^2$. 
In the LPA$''$ the effective action is thus defined by the effective potential $U_k(\rho)$ and two functions of $\p^2$, $Z_k(\p^2)$ and $Y_k(\p^2)$, which we simply denote by $Z_k(\p)$ and $Y_k(\p)$ in the following. The transverse and longitudinal parts of the two-point vertex~(\ref{Gamma2}) in a uniform field $\phibf$ are obtained from 
\beq
\begin{split}
\Gamma_{A,k}(\p,\rho) &= Z_k(\p)\p^2 + U'_k(\rho) , \\ 
\Gamma_{B,k}(\p,\rho) &= \frac{Y_k(\p)}{2}\p^2 + U''_k(\rho) .
\end{split}
\label{GamAB_lpapp}
\eeq
Thus the main improvement of the LPA$''$ over the LPA$'$ is that the full momentum dependence of the propagator $G_k(\p,\phibf)$ is preserved by virtue of the momentum dependence of $Z_k(\p)$ and $Y_k(\p)$. In the LPA$'$, where the momentum dependence of $Z_k(\p)\equiv Z_k$ and $Y_k(\p)\equiv Y_k$ is neglected, we obtain a $\p^2$ variation of the two-point vertex. This $\p^2$ dependence is valid for $|\p|\ll k$ (which corresponds to the domain of validity of DE and LPA$'$) and is due to the regulator term $\Delta S_k$ which ensures that all vertices are regular functions of $\p^2/k^2$ in the limit $\p/k\to 0$. The anomalous dimension $\eta$ can be computed since $Z_k$ diverges as $k^{-\eta}$ at the critical point but the LPA$'$ does not allow us to obtain the full momentum dependence of the propagator ({\it stricto sensu} the LPA$'$ is valid only for $\p\to 0$ in the limit $k\to 0$). In the LPA$''$, we expect 
\beq 
G_{k=0}(\p,\phibf=0) \sim |\p|^{-2+\eta}, \quad \mbox{i.e} \quad  
Z_{k=0}(\p) \sim |\p|^{-\eta} 
\label{Glpapp_qcp} 
\eeq
at the QCP.\footnote{The regulator term $\Delta S_k$ ensures that the two-point vertex is a regular function of $\p$ for $|\p|\ll k$, i.e. $\Gamma^{(2)}_k(\p,\phibf=0)\sim k^{-\eta}\p^2$ at the critical point.} This result should be valid for $|\p|$ smaller than the Ginzburg momentum scale $p_G\sim u_0/24\pi$.\cite{Rancon13a} Thus the anomalous dimension $\eta$ can be retrieved from the momentum dependence of $Z_k(\p)$. 

When $N\geq 2$ the excitation gap in the disordered phase turns out to be very well approximated by\cite{[{This is not true for $N=1$; see }]Rose16a}
\beq
\Delta = \xi^{-1} = \lim_{k\to 0} \sqrt{\frac{U_k'(\rho=0)}{Z_k(\p=0)}} ,
\label{Delta} 
\eeq
which follows from the expansion to $\calO(\p^2)$ of $\Gamma^{(2)}_{k,\rm T}(\p)=\Gamma_{A,k}(\p)$ [Eq.~(\ref{GamAB_lpapp})]. In Sec.~\ref{subsec_spectral_chi} we shall see that in the disordered phase the spectral function $\chi_{\rm L,T}''(\w)$ exhibits a sharp peak at the energy $\Delta$ defined by~(\ref{Delta}).  
On the other hand the stiffness, defined by~(\ref{rhosdef}), is obtained from 
\beq
\rhos = 2 Z_k(\p=0) \rho_0 .
\label{rhos_def1} 
\eeq 

The RG equations for $Z_k(\p)$ and $Y_k(\p)$ can be obtained by using
\beq
\begin{split}
Z_k(\p) &= \frac{\Gamma^{(2)}_{k,\rm T}(\p,\rho_{0,k}) - \Gamma^{(2)}_{k,\rm T}(\p=0,\rho_{0,k})}{\p^2} , \\ 
Z_k(\p) + \rho_{0,k} Y_k(\p) &=  \frac{\Gamma^{(2)}_{k,\rm L}(\p,\rho_{0,k}) - \Gamma^{(2)}_{k,\rm L}(\p=0,\rho_{0,k})}{\p^2} ,
\end{split}
\eeq
and the RG equation satisfied by $\Gamma_k^{(2)}(\p,\phibf)$ (see Appendix~\ref{app_rgeq}). In the usual way, we define $Z_k(\p)$ and $Y_k(\p)$ from the two-point vertex evaluated at the minimum $\rho_{0,k}$ of the effective potential $U_k(\rho)$.\cite{Berges02}

\subsubsection{Scalar susceptibility}

In the presence of a nonzero external source $h$, we consider the following ansatz for the effective action, 
\begin{align} 
\Gamma^{\mathrm{LPA}''}_k[\phibf,h] ={}& \Gamma^{\mathrm{LPA}''}_k[\phibf]
+ \half \int_\x h(\x) f_k(-\boldsymbol{\partial}^2) \phibf(\x)^2  \nonumber \\ & + \half \int_\x h(\x) \gamma_k(-\boldsymbol{\partial}^2) h(\x) ,
\label{lpapp1} 
\end{align} 
with initial conditions $f_\Lambda(\p)=-2$ and $\gamma_\Lambda(\p)=0$. In addition to the effective potential, the effective action includes four functions of momentum: $Z_k(\p)$, $Y_k(\p)$, $f_k(\p)$ and $\gamma_k(\p)$. Equation~(\ref{lpapp1}) yields 
\beq 
\begin{split} 
\Gamma^{(1,1)}_{i,k}(\p,\rho) &= \phi_i f_k(\p) , \\ 
\Gamma^{(0,2)}_{k}(\p,\rho) &= \gamma_k(\p) ,
\end{split}
\eeq 
in agreement with the general form of $\Gamma^{(1,1)}$ and $ \Gamma^{(0,2)}$ (Sec.~\ref{sec_nprg:subsubsec_chis}). The functions $f_k$ and $\gamma_k$ do not depend on $\rho$ in the LPA$''$. Their flow equations can be deduced from $\dk\Gamma^{(1,1)}_{i,k}$ and $\dk\Gamma^{(0,2)}_{k}$ (see Appendix~\ref{app_rgeq}).

\subsubsection{Conductivity}

In the presence of a nonzero gauge field $\A$ we make the effective action~(\ref{lpapp}) gauge invariant by replacing the derivative $\dmu$ by the covariant derivative $D_\mu$. $\Gamma_k[\phibf,\A]$ may also include terms depending on the field strength 
\beq
F_{\mu\nu} = - [D_\mu,D_\nu] = \dmu \Anu - \dnu \Amu - [\Amu,\Anu] . 
\label{Fmunu} 
\eeq
From $F_{\mu\nu}$ one can construct two invariant terms, namely $\tr(F^2_{\mu\nu})$ and $(F_{\mu\nu}\phibf)^2$.\cite{Rose17} We therefore consider the effective action 
\begin{multline}
\Gamma_k^{{\rm LPA}''}[\phibf,\A] = \int_\x \biggl\lbrace \half (D_\mu\phibf) \cdot Z_k(-\D^2) (D_\mu\phibf) \\ + \quarter (\partial_\mu \rho) Y_k(-\boldsymbol{\partial}^2) (\partial_\mu \rho)  
+ \dfrac{1}{4} F_{\mu \nu}^a X_{1,k}(-\D^2) F_{\mu \nu}^a  \\ + \dfrac{1}{4}  F_{\mu \nu}^a T^a\phibf \cdot X_{2,k}(-\D^2) F_{\mu \nu}^b T^b\phibf  +  U_k(\rho)  \biggr\rbrace 
\label{lpapp2} 
\end{multline}
with initial conditions $X_{1,\Lamb}=X_{2,\Lamb}=0$. 
Since the covariant derivative of a scalar is equal to its regular derivative, $Y_k$ is a function of $-\boldsymbol{\partial}^2$. The commutator $[\Amu,\Anu]$ in~(\ref{Fmunu}) contributes to the effective action to order $\calO(\Amu^3)$ and can be neglected when calculating the conductivity. In addition to the effective potential, the effective action now includes four functions of momentum: $Z_k(\p)$, $Y_k(\p)$, $X_{1,k}(\p)$ and $X_{2,k}(\p)$. In an LPA$'$-like approximation, one would simply neglect the momentum dependence of these functions. In the DE to second order they would be functions of $\rho$ rather than $\p$. As discussed in detail in Ref.~\onlinecite{Rose17} the DE runs into difficulties due to $X_{1,k}(\p)$ and $X_{2,k}(\p)$ being singular when $\p,k\to 0$ and $k/\p\to 0$; for instance it does not enable to compute the universal conductivity at the QCP. 

The functions $X_{1,k}(\p)$ and $X_{2,k}(\p)$ fully determine the vertices $\Gamma^{(1,1)a}_{j\mu}$ and $\Gamma^{(0,2)ab}_{\mu\nu}(\p,\phibf)$: 
\beq 
\Gamma^{(1,1)a}_{j\mu}(\p,\phibf) = ip_\mu (T^a\phibf)_j Z_k(\p) , 
\eeq
and
\begin{multline} 
\Gamzdmunuab(\p,\phibf) = p_\mu p_\nu \{ - \delta_{ab} X_{1,k}(\p) \\ - (T^a\phibf)\cdot(T^b\phibf) X_{2,k}(\p) \} 
+ \delta_{\mu\nu} \{ \delta_{ab} \p^2 X_{1,k}(\p) \\ + (T^a\phibf)\cdot(T^b\phibf) [Z_k(\p) + \p^2 X_{2,k}(\p) ] \} .
\label{gamdz} 
\end{multline}
Comparing with~(\ref{vertA}), we find 
\beq
\begin{split}
& Z_k(\p) = \Psi_{A,k}(\p), \quad 
X_{1,k}(\p) = -\Psi_{B,k}(\p), \\  
& X_{2,k}(\p) = -\Psi_{C,k}(\p), \quad 
\p^2 X_{1,k}(\p) = \bar\Psi_{B,k}(\p), \\
& Z_k(\p)+\p^2 X_{2,k}(\p) = \bar\Psi_{C,k}(\p)
\end{split}
\eeq 
in agreement with the Ward identities.\cite{Rose17} The various functions $\Psi$ and $\bar\Psi$ do not depend on $\rho$ in the LPA$''$. RG equations for $X_{1,k}(\p)$ and $X_{2,k}(\p)$ can therefore be derived from $\dk\Gamma^{(0,2)ab}_{\mu\nu}(\p,\phibf)$ (see Appendix~\ref{app_subsec_floweq}). 

Using~(\ref{gamdz}) we finally obtain 
\begin{multline}
K(i\wn) = - \delta_{\mu\nu} \{ \delta_{ab} \wn^2 X_{1,k}(i\wn) \\ + 
(T^a\bar\phibf)\cdot (T^b\bar\phibf) [ Z_k(i\wn) + \wn^2 X_{2,k}(i\wn) ] \} 
\end{multline}
in the quantum model. This yields 
\beq
\sig_k(i\wn) = 2\pi\sigQ \wn X_{1,k}(i\wn)  
\label{sigdis}
\eeq
in the disordered phase, and 
\beq
\begin{split}
\sigAk(i\wn) ={}& 2\pi\sigQ \biggl\{ \frac{2\rho_{0,k}}{\wn} Z_k(i\wn) \\ & +
\wn  [ X_{1,k}(i\wn) + 2\rho_{0,k} X_{2,k}(i\wn) ] \biggr\}  , \\ 
\sigBk(i\wn) ={}& 2\pi\sigQ \wn X_{1,k}(i\wn) 
\end{split}
\label{sigord} 
\eeq 
in the broken-symmetry phase, where $\sigQ=q^2/h$ is the quantum of conductance and we have restored $\hbar$.

\subsection{Large-$N$ limit} 
\label{sec_NPRG:subsec_largeN} 

Both DE\cite{Berges02} and BMW\cite{Blaizot06,Rose15} approximation schemes are exact in the large-$N$ limit. A crucial ingredient in the derivation of this result is that the vertices, e.g. $\Gamma_{A,k}(\p,\rho)$ and $\Gamma_{B,k}(\p,\rho)$, are field dependent. Since $Z_k(\p)$, $Y_k(\p)$, etc. are field independent in the LPA$''$, we do not expect the latter to be exact in the large-$N$ limit. However, it is possible to show that i) the potential $U(\rho)$ as well as the two-point correlation functions $\chi_{\rm L,T}(\p)$ [Eq.~(\ref{chi_def})] are correctly determined in the large-$N$ limit; and ii) the LPA$''$ is exact in the large-$N$ limit in the ordered phase, including the QCP.

To prove the above claims, we examine the flow equations (provided in Appendix~\ref{app_rgeq}) in the large-$N$ limit. The proof follows closely what is done in Refs.~\onlinecite{Blaizot06,Rose15} to solve the (similar) BMW equations in the large-$N$ limit. As the action $\Gamma_k[\phibf]$ and the field $\phibf$ respectively scale like $N$ and $\sqrt{N}$ one deduces that $W_k(\rho)=U'_k(\rho)$, $Z_k(\p)$ and the propagators $\GLTk(\p,\rho)$ are $\Oc(1)$ while $Y_k(\p)$ is $\Oc(1/N)$. 
Thus
\begin{equation}
\partial_k Z_k(\p) = \Oc(1/N)
\end{equation}
and $Z_k(\p)=1+\Oc(1/N)$. The large-$N$ transverse propagator reads
\begin{equation}
G_{k,\rm T}^{-1}(\q,\rho)=\q^2+W_k(\rho)+R_k(\q) +\Oc\left (\dfrac{1}{N}\right )
\end{equation} 
and the flow equation of $W_k(\rho)$ reduces to 
\begin{equation}
\partial_k W_k(\rho)  =  \frac{N}{2} W'_k(\rho) \tilde\partial_k  \int_\qv G_{k,\rm T}(\q,\rho)+\Oc\left (\dfrac{1}{N}\right )
\label{dWk}
\end{equation}
where $\tilde\partial_k=(\dk R_k)\partial_{R_k}$ acts only on the $k$ dependence of the cutoff function $R_k$. Equation~(\ref{dWk}) can be integrated using the change of variables $(k,\rho)\to (k,W)$ to yield the correct large-$N$ potential.\cite{Blaizot06} This also proves that the transverse propagator is exactly determined.

\begin{figure}[t!] 
\centering
\includegraphics{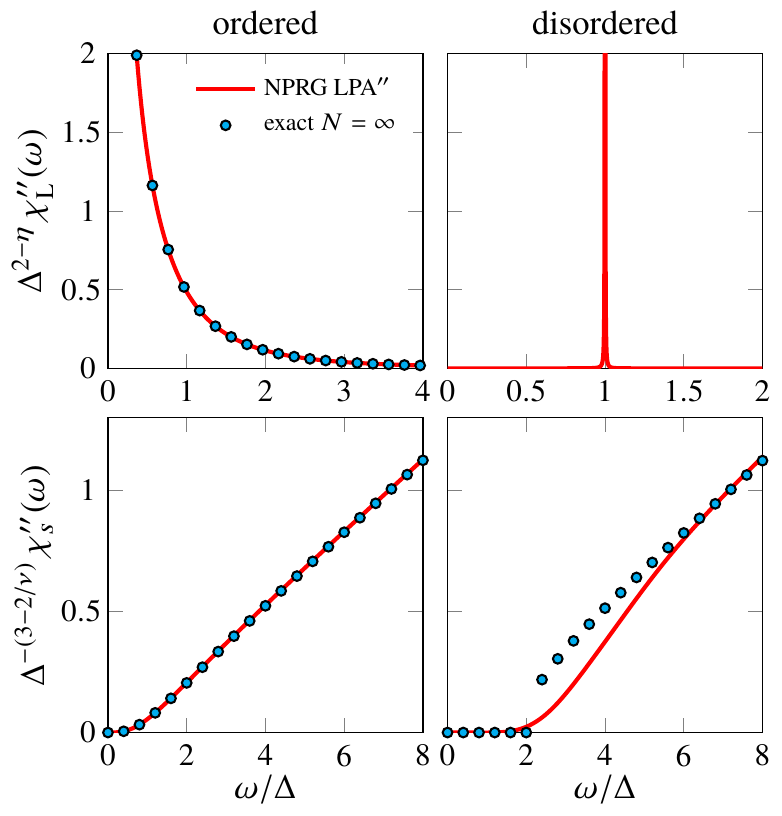}
\caption{(Top) Spectral function $\chiL''(\omega)$ in the ordered and disordered phases for $N=1000$ (solid line), compared to the exact large-$N$ solution (symbols). In the disordered phase, the exact solution for $\chiL''(\omega)=\chiT''(\omega)\sim \delta(\omega-\Delta)$ is not shown. 
(Bottom) Same as top panel but for the spectral function $\chis''(\omega)$.}
\label{fig_largeN_chi} 
\end{figure}

We now turn to $Y_k(\p)$, or equivalently to $\Gamma_{B,k}(\p)=\Gamma_{B,k}(\p,\rho_{0,k})$ [Eq.~(\ref{Gamma2})]. 
One has
\begin{align}
\partial_k \Gamma_{B,k}(\p) ={}& -\dfrac{N}{2} \Gamma_{B,k}(\p)^2  \tilde\partial_k \int_\qv G_{k,\rm T}(\q,\rho_{0,k}) \nnl
&\times G_{k,\rm T}(\p+\q,\rho_{0,k}) + \Oc\left (\dfrac{1}{N^2}\right ).\label{eq:gamBlargeN}
\end{align}
This agrees with the BMW equation in the case where $\Gamma_{B,k}$ does not depend on $\rho$.\cite{Rose15} 
Because of this lack of $\rho$ dependence, the change of variables $(k,\rho)\to (k,W)$ performed in the BMW equations is not possible. That difference is crucial: numerical integration of the flow equations shows that in the disordered phase $\Gamma_{B,k}(\p)$ differs from its exact value. However, in the ordered phase and at the critical point, one remarks that since $W_{k}(\rho_{0,k})=0$ for all $k$, $\tilde \partial_k G_{k,\rm T}(\q,\rho_{0,k})=\partial_k G_{k,\rm T}(\q,\rho_{0,k})$ and the rhs of \eqref{eq:gamBlargeN} becomes a total derivative. Integrating \eqref{eq:gamBlargeN} then yields the exact result. Since $Y_k(\p)$ only contributes to the longitudinal propagator in the ordered phase this means that $\chiL$ is exactly determined in the whole phase diagram, as evidenced in Fig.~\ref{fig_largeN_chi}~(top) where we compare $\chiL''(\w)$ to the exact result in the limit $N\to\infty$.

A similar analysis can be performed for the two functions $f_{k}(\p)$ and  $\gamma_{k}(\p)$ intervening in the scalar susceptibility. In the ordered phase the exact solution is recovered while in the disordered phase the LPA$''$ does not yield the exact result.
This is illustrated in Fig.~\ref{fig_largeN_chi}~(bottom) where we show the spectral function $\chis''(\w)$ for $N=1000$ as well as the exact result in the limit $N\to\infty$. 

In the disordered phase the solution of the RG flow for the conductivity differs from the exact solution.\cite{Rose17a} In the ordered phase, to leading order in $1/N$, $\sigA(\w)$ is determined by $\rho_{0,k}$ and $Z_k(\p)$, which reproduce the exact solution in the large-$N$ limit. No simple analytic form has been found for the next-to-leading order contribution to $\sigA(\w)$ which depends on $X_{1,k}$ and $X_{2,k}$. $\sigB(\w)$ is determined by the function $X_{1,k}(\p)$. For $\p=0$, it is possible to integrate the flow equation of $X_{1,k}(\p=0)$, following what is done in Ref.~\onlinecite{Rose17} to integrate the flow of $X_{1,k}(\rho)$ within DE, which yields the exact solution. At finite momentum, no analytic way to integrate the flow equation of $X_{1,k}(\p)$ has been found but the numerical integration of the flow equations shows an agreement with the exact solution up to numerical error.

\section{Spectral functions}
\label{sec_spectral}

The flow equations are given in Appendix~\ref{app_rgeq}. They can be solved in the usual way (see, e.g., Ref.~\onlinecite{Rose17}). Since the QCP manifests itself as a fixed point of the RG equations if we use dimensionless variables, we express all quantities in unit of the running scale $k$ (see Appendix~\ref{app_adim}). The flow equations are solved numerically for several sets of initial conditions $(r_0,u_0)$.
For a given value of $u_0$, the QCP can be reached by fine tuning $r_0$ to its critical value $r_{0c}$. We use $u_0\sim 50$ and $\Lamb=1$. 
The universal regime near the QCP can then be studied by tuning $r_0$ slightly away from $r_{c0}$. Universality of the results can be checked by changing the value of $u_0$ and the various correlation functions can be written in terms of universal scaling functions. Below we first discuss the two-point correlation function before turning to the scalar susceptibility and conductivity. We consider only the cases $N=2$ and $N=3$. When considering the two-point correlation function and scalar susceptibility we take the freedom to adjust the (nonuniversal) scale of correlation functions and spectral functions.

\begin{widetext} 

\begin{table}
\centering
\caption{Critical exponent $\nu$ for the three-dimensional classical O($N$) universality class obtained in the NPRG approach, from DE, LPA$''$ and  BMW approximation (results from the authors), compared to Monte Carlo (MC) simulations, (perturbative) field theories (FT) and conformal bootstrap (CB). The exponent has been determined before within NPRG for most values of $N$, see e.g. Ref.~\onlinecite{Gersdorff01} for DE, \onlinecite{Hasselmann12} for LPA$''$ and \onlinecite{Benitez09,Benitez12} for BMW approximation. For coherence we give our own results,  which are in agreement with the above references. For the BMW scheme we used the exponential regulator~(\ref{Rkdef}) with a parameter value $\alpha=2.25$ while for DE and LPA$''$ the values are obtained by applying the `principle of minimum sensitivity', i.e., by finding a value of $\alpha$  that extremizes $\nu$ (in practice $\alpha$ is close to $2$).}
\begin{tabular}{lllllll}
\hline \hline
\multicolumn{1}{c}{$N$}         & \multicolumn{1}{c}{DE} & \multicolumn{1}{c}{LPA$''$} & \multicolumn{1}{c}{BMW} & \multicolumn{1}{c}{MC} & \multicolumn{1}{c}{FT} & \multicolumn{1}{c}{CB} \\
\hline
$1$        & $0.638$        &    $0.631$    & $0.632$    & $0.63002(10)$\hfill~\cite{Hasenbusch10}& $0.6306(5)$~\hfill\cite{Pogorelov08}    & $0.629971(4)$\cite{Kos16} \\
$2$     & $0.668$    &    $0.679$    & $0.673$ & $0.6717(1)$\cite{Campostrini06} & $0.6700(6)$\cite{Pogorelov08}    & $0.67191(12)$\cite{Kos16} \\
$3$     & $0.706$    &    $0.725$    & $0.714$ & $0.7112(5)$\cite{Campostrini02} & $0.7060(7)$\cite{Pogorelov08}    & $0.7121(28)$\cite{Kos16} \\
$4$     & $0.741$    &    $0.765$    & $0.754$ & $0.749(2)$\cite{Hasenbusch01}     & $0.741(6)$\cite{Guida98} & \\
$5$        & $0.774$    &    $0.799$    & $0.787$ &       & $0.766$\cite{Antonenko95}& \\
$6$        & $0.803$    &    $0.836$ & $0.816$ &       & $0.790$\cite{Antonenko95}& \\
$8$        & $0.848$    &    $0.866$    & $0.860$ &       & $0.830$\cite{Antonenko95}& \\
$10$     & $0.879$    &    $0.892$    & $0.893$ &       & $0.859$\cite{Antonenko95}& \\
$100$     & $0.989$    &    $0.990$    & $0.990$ &       & $0.989$\cite{Moshe03}    & \\
$1000$    & $0.999$    &    $0.999$    & $0.999$ &       & $0.999$\cite{Moshe03}     & \\
\hline
\end{tabular}
\label{table_nu}
\caption{Same as~\ref{table_nu} but for the anomalous dimension $\eta$.}
\begin{tabular}{lllllll}
\hline \hline
\multicolumn{1}{c}{$N$}         & \multicolumn{1}{c}{DE} & \multicolumn{1}{c}{LPA$''$} & \multicolumn{1}{c}{BMW} & \multicolumn{1}{c}{MC} & \multicolumn{1}{c}{FT} & \multicolumn{1}{c}{CB} \\
\hline
$1$        & $0.0443$        & $0.0506$        & $0.0411$        & $0.03627(10)$\cite{Hasenbusch10}& $0.0318(3)$\cite{Pogorelov08}& $0.036298(2)$\cite{Kos16}  \\
$2$        & $0.0467$ & $0.0491$ & $0.0423$ & $0.0381(2)$\cite{Campostrini06} & $0.0334(2)$\cite{Pogorelov08}& $0.03852(64)$\cite{Kos16} \\
$3$        & $0.0463$ & $0.0459$ & $0.0411$ & $0.0375(5)$\cite{Campostrini02} & $0.0333(3)$\cite{Pogorelov08}& $0.0385(11)$\cite{Kos16}\\
$4$        & $0.0443$ & $0.0420$ & $0.0386$ & $0.0365(10)$\cite{Hasenbusch01} & $0.0350(45)$\cite{Guida98} & \\
$5$        & $0.0413$ & $0.0382$  & $0.0354$ &   & $0.034$\cite{Antonenko95}     & \\
$6$        & $0.0381$ & $0.0346$    & $0.0321$ &   & $0.031$\cite{Antonenko95}    &  \\
$8$        & $0.0319$    & $0.0287$    & $0.0264$     &      & $0.027$\cite{Antonenko95} &   \\
$10$    & $0.0270$    & $0.0243$    & $0.0220$ &   & $0.024$\cite{Antonenko95}     &  \\
$100$    & $0.00296$    & $0.00289$ & $0.00233$ &      & $0.0027$\cite{Moshe03}        & \\
$1000$    & $0.000296$& $0.000293$& $0.000233$ &     & $0.00027$\cite{Moshe03}        & \\
\hline
\end{tabular}
\label{table_eta}
\end{table}
\end{widetext}

\begin{figure*}
\parbox{8.5cm}{
\centering
\includegraphics{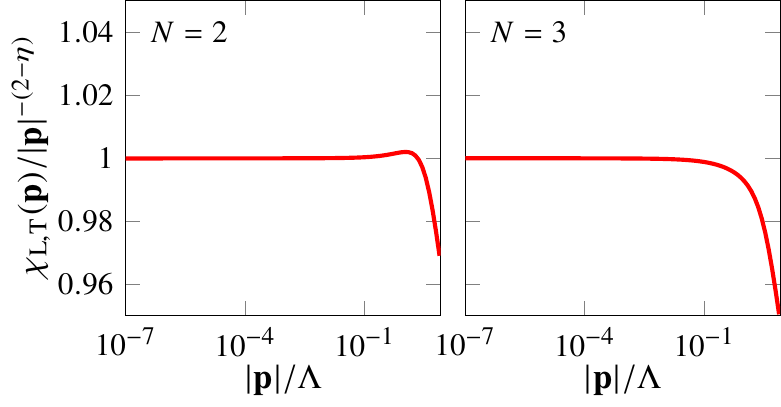}
\caption{Two-point correlation function $\chi_{\rm L,T}(\p)/|\p|^{-2+\eta}$ at the QCP for $N=2$ and $N=3$. The normalization is chosen to have a ratio equal to one for $\p\to 0$.}
\label{fig_chiLT_QCP}
\centering
\includegraphics{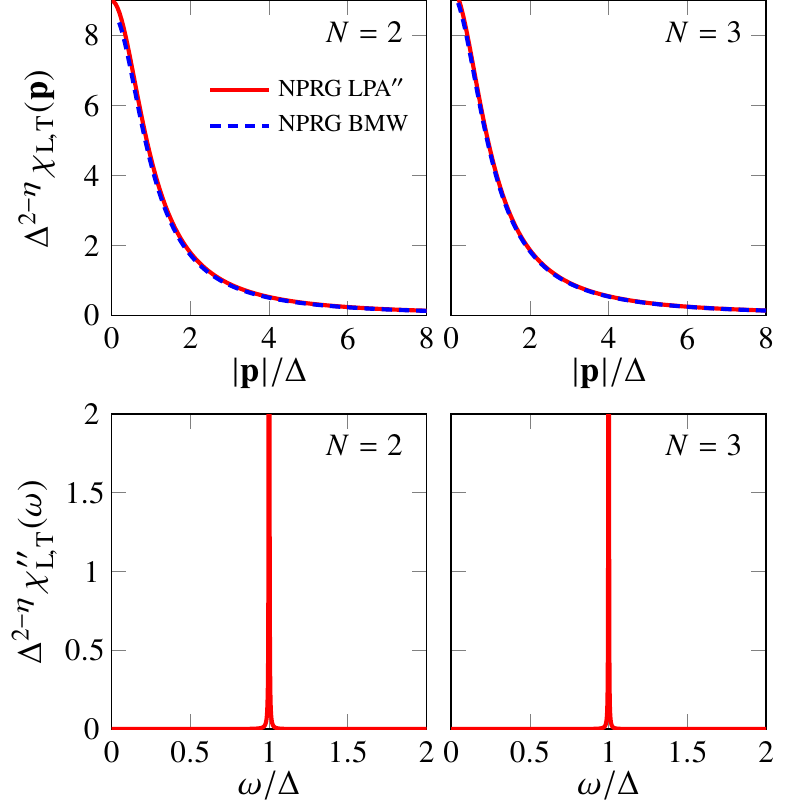}
\caption{$\chiLT(\p)$ and $\chiLT''(\w)$ in the disordered phase for $N=2$  and $N=3$.}
\label{fig_deso_compa_chiLT}}
\hspace{0.5cm}
\parbox{8.5cm}{
\centering
\includegraphics{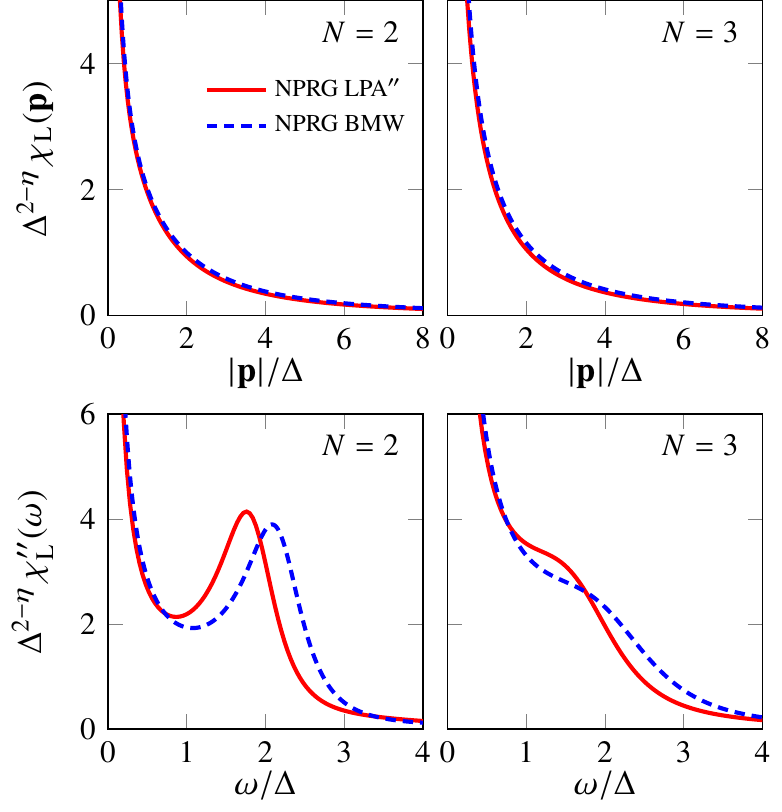}
\caption{$\chiL(\p)$ and $\chiL''(\w)$ in the ordered phase for $N=2$ and $N=3$. }
\label{fig_ordo_compa_chil}
\centering
\includegraphics{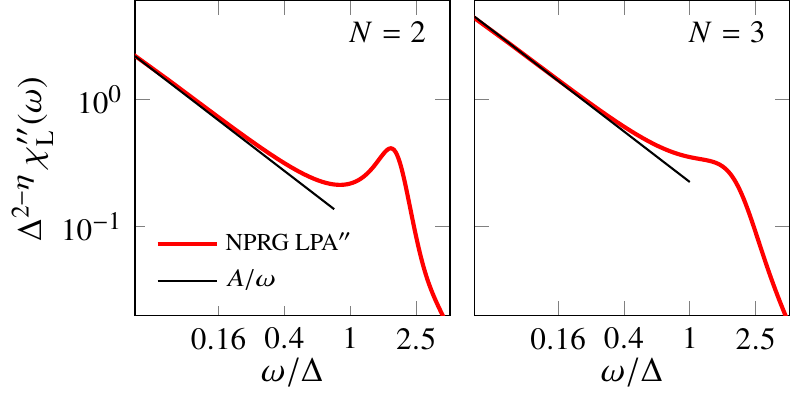}
\caption{Log-log scale plot of $\chiL''(\w)$ in the ordered phase for $N=2$ and $N=3$, showing the asymptotic behavior $\chiL''(\w)\sim 1/\w$ at low energies.}
\label{fig_ordo_chil_loglog}}
\end{figure*}

\subsection{Two-point correlation function} 
\label{subsec_spectral_chi} 

In the universal regime near the QCP, the two-point correlation function $\chi_{\alpha=\rm L,T}$ [Eq.~(\ref{chi_def})] and its spectral function satisfy the scaling forms\cite{Sachdev_book}
\begin{equation}
\begin{split}
\chi_\alpha(\p) &= Z_{\alpha,\pm} \Delta^{\eta-2} \tilde\Phi_{\alpha,\pm} \left( \frac{|\p|}{\Delta} \right) , \\
\chi''_\alpha(\w) & = {\rm Im}[\chi^R_\alpha(\w)] = Z_{\alpha,\pm} \Delta^{\eta-2} \Phi_{\alpha,\pm} \left( \frac{\w}{\Delta} \right) ,
\end{split}
\label{chiLT_scaling}
\end{equation}
where $\eta$ is the anomalous dimension of the $\varphibf$ field at the QCP. Recall that $\chi^R(\w)=\chi(|\p|\to -i\w+0^+)$ is the retarded susceptibility. $\tilde\Phi_{\alpha,\pm}$ and $\Phi_{\alpha,\pm}$ are universal scaling functions and $Z_{\alpha,\pm}$ a nonuniversal constant with dimension of (length)$^\eta$. The index $+/-$ refers to the disordered and ordered phases, respectively. $\Delta$ is a characteristic energy scale given by the excitation gap in the disordered phase. In the ordered phase, we take $\Delta$ to be the excitation gap in the disordered phase at the point located symmetrically wrt the QCP (i.e. corresponding to the same value of $|r_0-r_{0c}|$). Since $\Delta$ vanishes at the QCP, Eqs.~(\ref{chiLT_scaling}) imply $\chi_\alpha(\p)\sim |\p|^{\eta-2}$ and $\chi''_\alpha(\w)\sim|\w|^{\eta-2}$ when $r_0=r_{0c}$. Since $\chi''_{\rm L}(\w)$ and $\chi''_{\rm T}(\w)$ are odd in $\w$ we shall only consider the case $\w\geq 0$ in the following.

\subsubsection{QCP}
\label{subsubsec_spectral_chi_qcp}

At the QCP the anomalous dimension $\eta$ is given by the value of the running anomalous dimension $\eta_{k}=-k\dk\ln Z_k(\p=0)$ reached when $k\to 0$. The correlation-length-exponent $\nu$ can be obtained from the runaway flow from the fixed point when the system is not exactly at criticality (which, in practice, is always the case), e.g., $\trho_{0,k}\simeq \trho^*_0+ \const\times (\Lambda/k)^{1/\nu}$ ($\trho_0$ is the dimensionless field variable, see Appendix~\ref{app_adim}, and $\trho^*_0$ its fixed-point value). Results  obtained for various values of $N$ are shown in Tables~\ref{table_nu} and \ref{table_eta} where we compare the LPA$''$ to other methods. The LPA$''$ provides us with satisfying values for the critical exponent $\nu$ (within 2\% of the conformal bootstrap results for $N=1,2,3$) but is less accurate, and significantly less reliable than the DE and BMW approximations, for the anomalous dimension. 

We thus conclude that, when improving the approximation scheme starting from the LPA$'$, it is more efficient to include the full field dependence (as in DE) than the full momentum dependence (as in LPA$''$) of $Z_k$ and $Y_k$. Naive power counting near four dimensions shows that indeed the field dependence is more important than the momentum dependence so that, at least near four dimensions, the superiority of DE over LPA$''$ in estimating the anomalous dimension should not come as a surprise. We note however that any field truncation in DE is likely to strongly deteriorate the estimate of $\eta$ below the accuracy of LPA$''$.\cite{[{See, e.g., Table VIII in }]Delamotte04} It is therefore natural to ascribe the lack of accuracy of the LPA$''$ to the neglect of diagrams involving the momentum dependence of $\Gamma^{(3)}$ or $\Gamma^{(4)}$.\footnote{Indeed the neglect of these diagrams appears as the main difference between LPA$''$ and BMW, the latter giving a much better estimate of the anomalous dimension. See, e.g., Ref.~\onlinecite{Papenbrock95} for a discussion of these diagrams.} In any case the anomalous dimension is small for the three-dimensional O($N$) model and an accurate estimate is not crucial when focusing on the full momentum dependence (which is not dominated by $\eta$ on a typical scale fixed by $\Delta$). 

The momentum dependence of $\chi_{\rm L,T}(\p)$ at criticality is shown in Fig.~\ref{fig_chiLT_QCP}. At small momentum, below the Ginzburg momentum scale $p_G\sim u_0/24\pi\sim \Lambda/2$, $\chi_{\rm L,T}(\p)\sim |\p|^{-2+\eta}$ in agreement with the expected result~(\ref{Glpapp_qcp}). The value of the exponent $\eta$ is the same as that obtained from the running anomalous dimension $\eta_k$.

\begin{figure*}
\parbox{8.5cm}{ 
\centering
\includegraphics{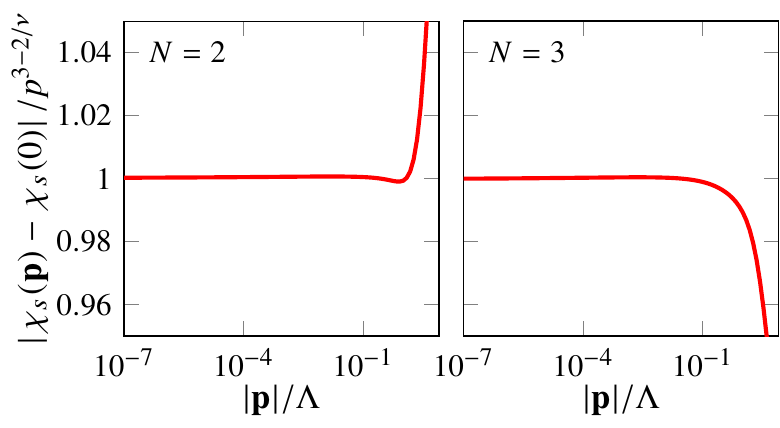}
\caption{$|\chis(\p)-\chis(0)|/|\p|^{3-2/\nu}$ at the QCP for $N=2$ and $N=3$. The normalization is chosen to have a ratio equal to one for $\p\to 0$.}
\label{fig_chis_QCP} 
\centering
\includegraphics{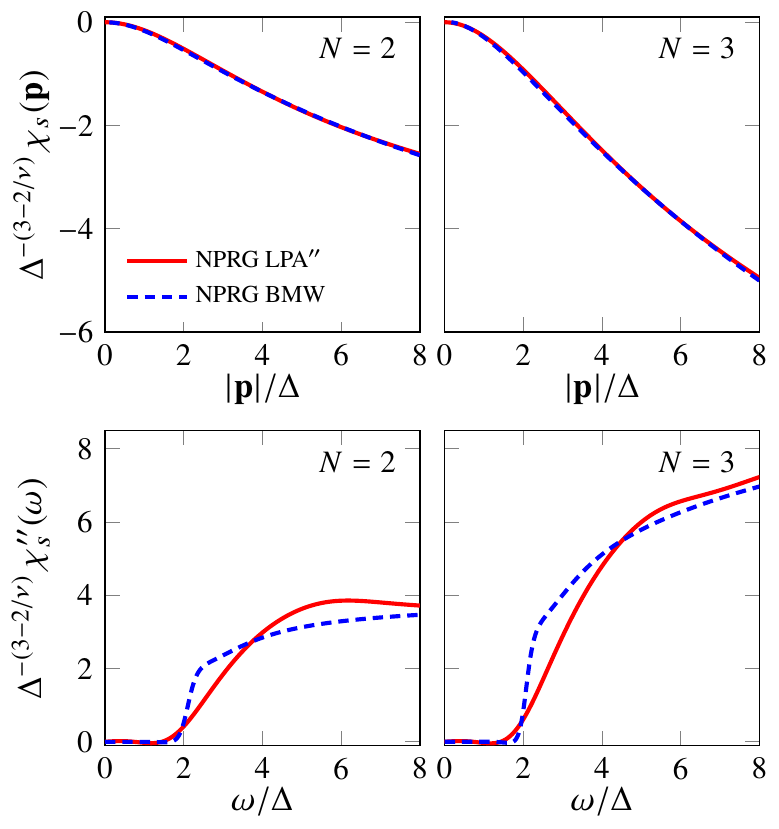}
\caption{$\chis(\p)$ and $\chis''(\w)$ in the disordered phase for $N=2$  and $N=3$.}
\label{fig_deso_compa_chis}}
\hspace{0.5cm}
\parbox{8.5cm}{ 
\centering
\includegraphics{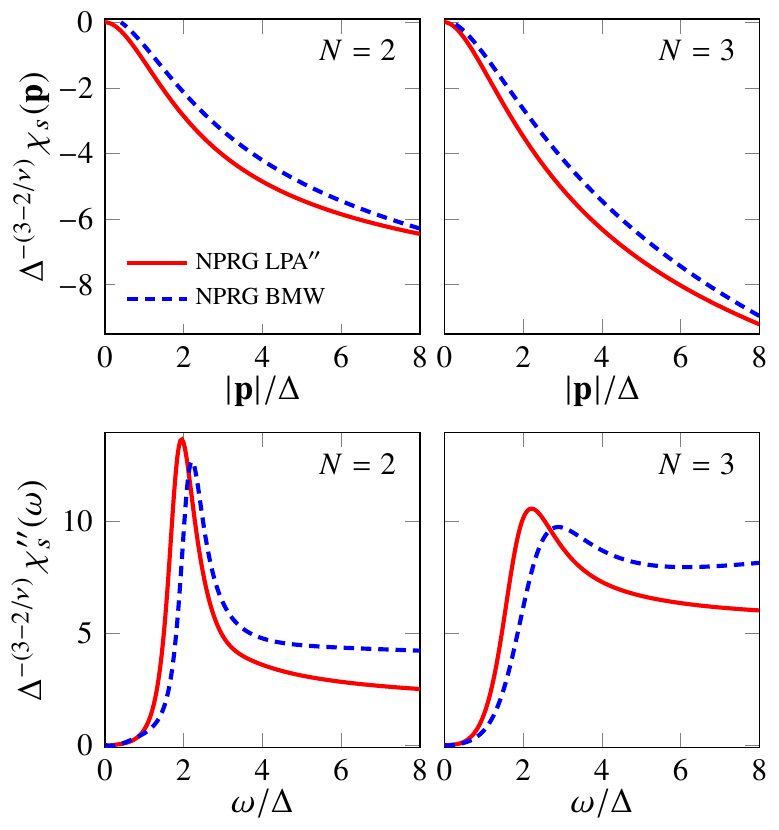}
\caption{$\chis(\p)$ and $\chis''(\w)$ in the ordered phase for $N=2$ and $N=3$. }
\label{fig_ordo_compa_chis} 
\centering
\includegraphics{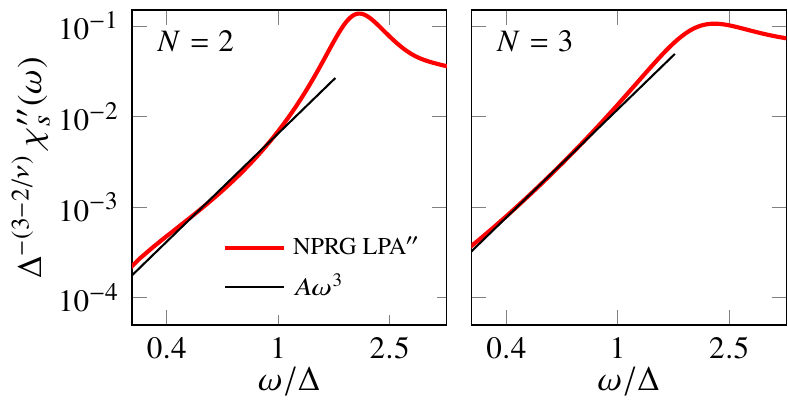}
\caption{Log-log scale plot of $\chis''(\w)$ in the ordered phase for $N=2$ and $N=3$, showing the asymptotic behavior $\chis''(\w)\sim \w^3$ at low energies.}
\label{fig_ordo_chis_loglog}} 
\end{figure*} 

\subsubsection{Disordered phase}

The two-point correlation function $\chi_{\rm L,T}(\p)$ in the disordered phase is shown in Fig.~\ref{fig_deso_compa_chiLT} for $N=2$ and $N=3$. (More precisely we show the universal scaling function $\tilde\Phi_{\alpha,+}(|\p|/\Delta)$.) The spectral function $\chi''_{\rm L,T}(\w)$, obtained from a numerical analytic continuation using Pad\'e approximants, consists of a narrow peak at an energy which is very well approximated by~(\ref{Delta}). The results are in very good agreement with the BMW results from Ref.~\onlinecite{Rose15}. Recall that when comparing LPA$''$ and BMW we take the freedom to adjust the (nonuniversal) relative scale.

\begin{table}[b]
\caption{Universal ratio $\rhos/N\Delta$ obtained in the NPRG approach from DE, LPA$''$ and BMW (results from the authors) compared to Monte Carlo (MC) simulations and exact diagonalization (ED). The exact result in the large-$N$ limit is $1/4\pi\simeq 0.0796$.}
\begin{tabular}{llllll}
\hline \hline
\multicolumn{1}{c}{$N$}         & \multicolumn{1}{c}{DE} & \multicolumn{1}{c}{LPA$''$} & \multicolumn{1}{c}{BMW} & \multicolumn{1}{c}{MC}    & \multicolumn{1}{c}{ED}   \\
\hline
$2$     & $0.207$    &    $0.195$    & $0.193$ & $0.220$\hfill~\cite{Gazit13a} & $0.17(2)$\hfill~\cite{Nishiyama17}  \\
$3$     & $0.147$    &    $0.140$    & $0.137$ & $0.114$\hfill~\cite{Gazit13a} &  \\
$4$     & $0.118$    &    $0.115$    & $0.111$ &    &       \\
$6$        & $0.0935$    &    $0.0947$    & $0.0903$     &    &        \\
$8$        & $0.0846$    &    $0.0876$    & $0.0829$    &    &        \\
$10$     & $0.0810$    &    $0.0844$    & $0.0803$ &    &        \\
$1000$    & $0.0795$    &    $0.0798$    & $0.0796$ &    &        \\
\hline
\end{tabular}
\label{table_rhos}
\end{table}

\subsubsection{Ordered phase}

The ordered phase is characterized by the stiffness $\rhos$ [Eq.~(\ref{rhos_def1})]. The ratio $\rhos/N\Delta$, where $\Delta$ is the excitation gap in the disordered phase at the point located symmetrically wrt the QCP (i.e. corresponding to the same value of $|r_0-r_{0c}|$) is a universal number equal to $1/4\pi$ in the large-$N$ limit. When $2\leq N\leq 4$, the LPA$''$ value of this ratio is between the results obtained from DE and BMW approximation, and in reasonable agreement with Monte Carlo simulations for $N=2$ and $3$ (Table~\ref{table_rhos}). For $N>4$ the LPA$''$ starts to deviate from DE and BMW but is nevertheless exact in the large-$N$ limit.

The longitudinal correlation function $\chi_{\rm L}(\p)$ in the ordered phase is shown in Fig.~\ref{fig_ordo_compa_chil} for $N=2$ and $N=3$. Again there is a very good agreement with the BMW result from Ref.~\onlinecite{Rose15}. The spectral function $\chi''_{\rm L}(\w)$ is also very similar in the two approaches: it shows a $1/\w$ divergence at low energies due to the coupling of longitudinal fluctuations to transverse ones\cite{Patasinskij73,Sachdev99,Zwerger04,Dupuis11} (Fig.~\ref{fig_ordo_chil_loglog}) and a broad peak around $\w=\Delta$ for $N=2$ (for $N=3$ the peak has disappeared but a faint structure can still be seen), presumably due the Higgs mode.\cite{Rose15}

\subsection{Scalar susceptibility}

In the universal regime near the QCP,\cite{Podolsky12} 
\begin{equation}
\begin{split}
\chis(\p) &= \calB_\pm + \calA_\pm \Delta^{3-2/\nu} \tilde\Phi_{s,\pm}\left(\frac{|\p|}{\Delta}\right) , \\  
\chis''(\w) &= {\rm Im}[ \chis^R(\w) ] =   \calA_\pm \Delta^{3-2/\nu} \Phi_{s,\pm}\left(\frac{\w}{\Delta}\right) ,
\end{split}
\label{chis_scaling}
\end{equation}
where $\tilde\Phi_{s,\pm}$ and $\Phi_{s,\pm}$ are universal scaling functions and $\calA_\pm,\calB_\pm$ nonuniversal constants. At the QCP ($\Delta=0$), $\chis(\p)-\chis(0)\sim |\p|^{3-2/\nu}$ and $\chi''_s(\w)\sim |\w|^{3-2/\nu}$. Since $\chis''(\w)$ is an odd function of $\w$, we shall only consider the case $\w>0$ in the following.

\begin{table}[b]
\centering
\caption{Critical exponent $\nu$ for the three-dimensional classical O($N$) universality class obtained in the LPA$''$, from either $\rho_{0,k}$ (Sec. \ref{subsubsec_spectral_chi_qcp}) or $\chis$ (Sec. \ref{subsubsec_spectral_chis_qcp}), compared to conformal bootstrap (CB). }
\begin{tabular}{lllllll}
\hline \hline
\multicolumn{1}{c}{$N$}         & \multicolumn{1}{c}{from $\rho_{0,k}$} & \multicolumn{1}{c}{from $\chis$} &  \multicolumn{1}{c}{CB~\cite{Kos16}} \\
\hline
$2$    &    $0.679$	& $0.639$    & $0.67191(12)$ \\
$3$    &    $0.725$ & $0.682$   & $0.7121(28)$ \\
$4$    &    $0.765$ & $0.722$    \\
$5$    &    $0.799$ & $0.756$    \\
$6$    &    $0.836$ & $0.784$    \\
$8$    &    $0.866$ & $0.827$    \\
$10$   &    $0.892$ & $0.856$    \\
$100$  &    $0.990$ & $0.984$    \\
$1000$ &    $0.999$ & $0.998$   \\
\hline
\end{tabular}
\label{table_nu_chis}
\centering
\setlength{\tabcolsep}{8pt}
\caption{Universal conductivity $\sig^*/\sigQ$ at the QCP, obtained with a regulator parameter value $\alpha=2.25$ [Eq.~(\ref{Rkdef})], compared to results obtained from quantum Monte Carlo simulations\cite{Witczak14,Chen14,Gazit13a,Katz14,Gazit14} (QMC) and conformal bootstrap\cite{Kos15} (CB). The exact value for $N\to \infty$ is $\pi/8\simeq 0.3927$.}
\begin{tabular}{cccc}
\hline \hline
$N$ & NPRG & QMC & CB \\
\hline
2 	& 0.3218 & 0.355-0.361 & 0.3554(6) \\
3	& 0.3285 \\
4 	& 0.3350\\ 
10	& 0.3599 \\ 
1000& 0.3927 \\ 
\hline \hline
\end{tabular}
\label{table_sigQCP}
\end{table}

\subsubsection{QCP}
\label{subsubsec_spectral_chis_qcp}

The scalar susceptibility $\chis(\p)-\chis(0)\sim|\p|^{3-2/\nu}$ at criticality is shown in Fig.~\ref{fig_chis_QCP}. The momentum dependence provides us with an alternative computation of the critical exponent $\nu$ (Table~\ref{table_nu_chis}). The results are significantly less accurate than those obtained from $\rho_{0,k}$ (Sec.~\ref{subsec_spectral_chi}) but improve over the results of Ref.~\onlinecite{Rancon14}. 

\subsubsection{Disordered phase}

Figure~\ref{fig_deso_compa_chis} shows that the scalar susceptibility $\chis(\p)$ obtained in the LPA$''$ is in nearly perfect agreement with the BMW result.\cite{Rose15} Yet the spectral functions $\chis''(\w)$ differ, the energy gap $2\Delta$ being not as sharply defined in the LPA$''$. The difference reflects the difficulty to obtain a gapped spectral function $\chi''(\w)\propto\Theta(\w-2\Delta)$ with Pad\'e approximants.

\subsubsection{Ordered phase}

In the ordered phase, although the agreement between LPA$''$ and BMW for $\chis(\p)$ is not perfect, the LPA$''$ spectral function $\chis''(\w)$ compares fairly well with the BMW one (Fig.~\ref{fig_ordo_compa_chis}). In particular, for $N=2$ we clearly observe a Higgs resonance at $\w_H\simeq 1.95\Delta$, to be compared with $\w_H\simeq 2.2\Delta$ in the BMW approach.\cite{Rose15} For $N=3$ a broadened resonance around $\w_H\simeq2.2\Delta$ ($\w_H\simeq 2.7\Delta$ with BMW) is still visible (the resonance is suppressed for higher values of $N$). At low frequencies our results are compatible with the expected $\w^3$ behavior (Fig.~\ref{fig_ordo_chis_loglog}).

\subsection{Conductivity}

In the critical regime the conductivity tensor satisfies the scaling form \cite{Fisher90,Damle97}
\begin{equation}
\begin{split} 
\sig(\w) & = \sig_Q \Sig_+\left( \frac{\w}{\Delta}\right) , \\ 
\sig_{A,B}(\w)  & = \sig_Q \Sig_-^{A,B}\left( \frac{\w}{\Delta}\right) ,
\end{split}
\label{sigscaling}
\end{equation}
where $\Sig_\pm$ is a universal scaling function and $\sigQ$ the quantum of conductance. As the conductivity is dimensionless in two space dimensions there is no nonuniversal prefactor, unlike $\chiLT$ and $\chis$.

\subsubsection{QCP}

At the QCP, the universal scaling functions reach a nonzero limit $\Sig_\pm(\infty)$ and the ratio $\sig(\w=0)/\sig_Q=\sig^*/\sig_Q=\Sig_\pm(\infty)$ is universal,\cite{Fisher90} equal to $\pi/8$ in the large-$N$ limit.\cite{Sachdev_book} The LPA$''$ recovers the exact result in the large-$N$ limit. For $N=2$, it gives a value in reasonable agreement with (although 10\% smaller than) results from QMC\cite{Witczak14,Chen14,Gazit13a,Katz14,Gazit14} and conformal bootstrap\cite{Kos15} (Table~\ref{table_sigQCP}).

\subsubsection{Disordered phase}

\begin{figure}
\includegraphics{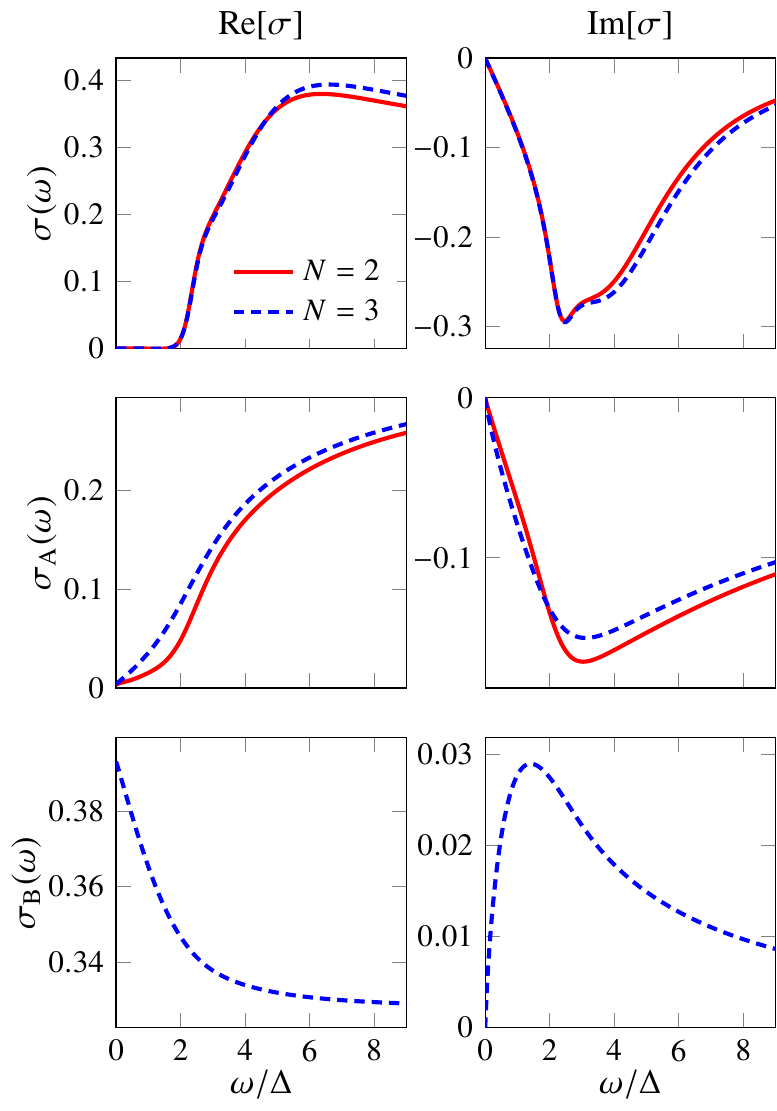}
\caption{
Conductivity $\sigma(\w)$ in the disordered and ordered phases for $N=2$ and $3$. The real and imaginary parts are respectively plotted left and right. Top: $\sigma(\w)$ in the disordered phase. Middle:  $\sigA(\w)$ in the ordered phase with the superfluid contribution subtracted. Bottom: $\sigB(\w)$ in the ordered phase. For $N=2$, $\sigB(\w)$ is not defined (see Sec.~\ref{sec_NPRG:subsubsec_cond}).
}
\label{fig_AllSig}
\end{figure}

The conductivity $\sigma(\w)$ in the disordered phase is shown in Fig.~\ref{fig_AllSig} (top panel). The system is insulating and the real part of $\sig(\w)$ vanishes below an energy gap $2\Delta$. The imaginary part varies linearly for $\w\ll\Delta$, i.e. $\sigma(\w)\simeq -i\Cdis\w$;  the system behaves as a perfect capacitor at low energies with capacitance (per unit area) $\Cdis=2\pi\hbar\sig_Q X_{1,k=0}(\wn=0)$. The ratio $\hbar\sig_Q/2\pi\Cdis\Delta$ is universal. The LPA$''$ value is in good agreement with the results of DE,\cite{Rose17} Monte Carlo simulations\cite{Gazit14} and exact diagonalization.\cite{Nishiyama17}

For large $N$, there is a discrepancy between the exact solution and our computation which has been noted in Sec.~\ref{sec_NPRG:subsec_largeN} for the two-point correlation function and the scalar susceptibility in the disordered phase. Furthermore the analytic continuation is made difficult by the singularity at $\w=2\Delta$ so that the frequency dependence of $\sig(\w)$ above $2\Delta$ should be taken with caution.  

\subsubsection{Ordered phase}

\begin{table}
\caption{Ratios $\hbar\sig_Q/2\pi \Cdis\Delta$ and $\Cdis/N\Lord\sig_Q^2$ obtained from the NPRG approach, compared to Monte Carlo (MC) simulations and exact diagonalization (ED). The exact results for $N\to\infty$ are  $6/\pi\simeq 1.90986$ and $1/24\simeq 0.041667$, respectively.}
\begin{tabular}{lllllll} \hline\hline
\multicolumn{1}{c}{$N$}
 & \multicolumn{4}{c}{$\hbar\sig_Q/2\pi \Cdis\Delta$} & \multicolumn{2}{c}{$\Cdis/N\Lord\sig_Q^2$} \\ \cmidrule(lr){2-5} \cmidrule(lr){6-7}
& \multicolumn{1}{c}{DE} & \multicolumn{1}{c}{LPA$''$} & \multicolumn{1}{c}{MC} & \multicolumn{1}{c}{ED} & \multicolumn{1}{c}{DE} & \multicolumn{1}{c}{LPA$''$} \\ \midrule
 $2$    & $1.98$    & $2.00$ & $2.1(1)$\hfill~\cite{Gazit14} & $2.0(4)$\hfill~\cite{Nishiyama17}    & $0.105$    & $0.0975$    \\
$3$        & $1.98$    & $1.98$    &    &    & $0.0742$    & $0.0706$    \\
$4$     & $1.98$    & $1.96$    &    &    & $0.0598$    & $0.0587$    \\
$5$     & $1.97$    & $1.94$    &    &    & $0.0520$    & $0.0526$    \\
$6$     & $1.97$    & $1.92$    &    &    & $0.0475$    & $0.0493$    \\
$8$     & $1.96$    & $1.90$    &    &    & $0.0431$    & $0.0461$    \\
$10$    & $1.96$    & $1.88$    &    &    & $0.0415$    & $0.0448$        \\
$100$    & $1.92$    & $1.80$    &    &    & $0.0413$    & $0.0443$    \\
$1000$    & $1.91$    & $1.79$    &    &    & $0.0416$    & $0.0446$    \\
\hline
\end{tabular}
\label{tab:cond_CoverL}
\end{table}

In the ordered phase, the conductivity tensor is defined by two independent elements, $\sigA(\w)$ and $\sigB(\w)$ [Eqs.~(\ref{sigord})]. The large-$N$ limit is exact. At low energies the system behaves as a superfluid or a perfect inductor, $\sigA(\w)\simeq i/\Lord(\w+i0^+)$, with inductance $\Lord=\hbar/2\pi\sig_Q\rhos$.  The ratio $\Cdis/N\Lord\sig^2_Q$ is universal (Table~\ref{tab:cond_CoverL}). 

$\sigA(\w)$, with the superfluid contribution $i/\Lord(\w+i0^+)$ subtracted, is shown in Fig.~\ref{fig_AllSig} (middle panel). Our results seem to indicate the absence of a constant $\calO(\wn^0)$ term in agreement with the predictions of perturbation theory.\cite{Podolsky11} Furthermore we see a marked difference in the low-frequency behavior of the real part of the conductivity between the cases $N=2$ and $N\neq 2$, but our numerical results are not precise enough to resolve the low-frequency power laws (predicted\cite{Podolsky11} to be $\w$ and $\w^5$ for $N\neq 2$ and $N=2$, respectively). 
On the other hand we  find that $\sigB(\w)$ reaches a nonzero universal value $\sigB^*$ in the limit $\w\to 0$ (Fig.~\ref{fig_AllSig}, bottom panel). Contrary to $\sig^*$, $\sigB^*$ turns out to be $N$ independent: the relative change in $\sigB^*$ is less than $10^{-6}$ when $N$ varies. Noting that the obtained value $\sigB^*/\sigQ\simeq 0.3927$ is equal to the large-$N$ result\cite{Rose17,Lucas17} $\pi/8$ within numerical precision, we have conjectured that $\sigB^*/\sigQ=\pi/8$ for all values of $N$.\cite{Rose17a}

\section{Conclusion}

We have presented an approximation scheme of the NPRG flow equations, the LPA$''$, that preserves the momentum dependence of correlation functions. As a test-bed we have considered the two-dimensional quantum O($N$) model. In the zero-temperature limit considered in the paper, this model is equivalent to the classical three-dimensional O($N$) model. Spectral functions of the two-dimensional quantum model can be obtained from an analytic continuation $|\p|\to -i\w+0^+$ using Pad\'e approximants. The LPA$''$ requires to solve coupled equations for the effective potential $U_k(\rho)$ and the momentum-dependent functions $Z_k(\p)$ and $Y_k(\p)$ that define the two-point vertex. To obtain the scalar susceptibility $\chis$ or the conductivity $\sigma$, additional equations for $f_k(\p)$ and $\gamma_k(\p)$ (for $\chis$), or $X_{1,k}(\p)$ and $X_{2,k}(\p)$ (for $\sigma$), must be considered. The fact that these functions depend only on momentum (and not also the field variable $\rho$) makes the approach relatively easy to implement numerically. 

We have made a detailed comparison of the results obtained within the LPA$''$ to those obtained from the DE or BMW approximation schemes. Overall the LPA$''$ remains relatively precise given its simplicity. The value of the critical exponent $\nu$ is nearly as accurate (at least for $N\leq 3$) as with DE or BMW but the anomalous dimension is less precise. As for the universal ratio $\rhos/\Delta$ between stiffness and excitation gap, the LPA$''$ result is very close to the BMW one. 
The universal scaling functions of various correlation functions (two-point correlation function, scalar susceptibility and conductivity) also compare satisfactorily  with the BMW results thus showing the ability of the LPA$''$ to reliably compute the momentum dependence. Indeed LPA$''$ and BMW show good qualitative agreement with some quantitative discrepancy. For instance in the LPA$''$ the Higgs resonance energy $\w_H/\Delta$ is equal to 1.95 and 2.2 for $N=2$ and 3, respectively, whereas BMW gives 2.2 and 2.7. A weakness of the LPA$''$ though is its inability to reproduce the large-$N$ limit in the disordered phase (i.e. when there is a gap in the spectrum).  

The LPA$''$ is particularly successful in computing the zero-temperature conductivity. In the presence of an external non-Abelian gauge field $\Amu$, it is not clear how to implement the BMW scheme in a gauge-invariant way. On the other hand DE breaks down at low energies due to some vertices being singular functions of momentum. In contrast the LPA$''$ allows us to obtain the full frequency dependence of the conductivity at the QCP and in the disordered and ordered phases. The value of the universal conductivity $\sig(\w\to 0)=\sig^*$ at the QCP is within 10\% of the conformal bootstrap result. An important result obtained by the LPA$''$ is the superuniversality of one of the elements of the conductivity tensor, $\sigB^*/\sig_Q=\pi/8$, in the ordered phase.\cite{Rose17a} 

Finally we would like to point out that the LPA$''$ might offer the possibility to avoid the analytic continuation of numerical data using Pad\'e approximants (or alternative methods). Indeed, by approximating the propagators in the internal loops of the flow equations by their LPA$'$ expressions, it becomes possible to perform exactly both Matsubara-frequency sums and analytic continuation to real frequencies,\cite{Kamikado14,Tripolt14,Tripolt14a,Wambach14,Pawlowski15,Cyrol18,Pawlowski17} which would allow to obtain the frequency dependence of correlation functions in the hydrodynamic regime $|\w|\lesssim T$.

\appendix

\section{RG equations in the LPA$''$}
\label{app_rgeq}

In this Appendix, we provide some technical details regarding the LPA$''$. Flow equations for the vertices are obtained by taking functional derivatives of Eqs.~(\ref{Weteq}), (\ref{Weteq_h}) and (\ref{Weteq_A}). Replacing the vertices by their LPA$''$ expressions, we derive equations for the various functions of interest: $W_k(\rho)=U'_k(\rho)$, $Z_k(\p)$, $Y_k(\p)$, $f_k(\p)$, $\gamma_k(\p)$, $X_{1,k}(\p)$ and $X_{2,k}(\p)$. 
To alleviate the notations in the following we do not write explicitly the $k$ index and $\rho$-dependence of the functions.

\begin{widetext} 
\subsection{Vertices}

In this section we list all vertices that enter the flow equations (besides those already considered in the text). We do not write the Kronecker symbol expressing the conservation of total momentum and set the volume equal to unity. All vertices are evaluated in a uniform field configuration and we use the notation $W'=\partial_\rho W$, etc. $\perm(1,\cdots,n)$ denotes all (different) terms obtained by permutation of $(\p_1,i_1;\cdots;\p_n,i_n)$. 

\subsubsection{Two-point correlation function}

The vertices entering the flow equation $\dk\Gamma^{(2)}$ are 
\begin{align}
\Gamma_{i_1 i_2 i_3}^{(3,0)}(\pv_1,\pv_2,\pv_3)  ={}&\dfrac{1}{2} Y(\pv_1)\p_1^2 \delta_{i_2i_3}\phi_{i_1} +W'  \delta_{i_1 i_2}\phi_{i_3} +W'' \phi_{i_1}\phi_{i_2}\phi_{i_3} + \perm(1,2,3), \\
\Gamma_{i_1i_2i_3i_4}^{(4,0)}(\pv_1,\pv_2,\pv_3,\pv_4)={}&\dfrac{1}{2}Y(\pv_1+\pv_2)(\pv_1+\pv_2)^2 \delta_{i_1 i_2}\delta_{i_3i_4} \nnl
&{}+{} W'\delta_{i_1i_2}\delta_{i_3i_4}{}+W'' \phi_{i_1}\phi_{i_2}\delta_{i_3i_4} +W''' \phi_{i_1}\phi_{i_2}\phi_{i_3}\phi_{i_4} + \perm(1,2,3,4).
  \label{vert3}
\end{align}

\subsubsection{Scalar susceptibility}

In the calculation of the scalar susceptibility, 
\begin{align}
\Gamma_{i_1 i_2}^{(2,1)}(\pv_1,\pv_2,\pv_3) &= \delta_{i_1 i_2} f(\pv_3), \\ 
\Gamma^{(3,1)}_{i_1 i_2 i_3}(\pv_1,\pv_2,\pv_3,\pv_4) &= \Gamma^{(2,2)}_{i_1 i_2}(\pv_1,\pv_2,\pv_3,\pv_4)=0.
\end{align}

\subsubsection{Conductivity}

In the calculation of the conductivity, 
\begin{align}
    \Gamma_{i_1 i_2,\mu}^{(2,1)a}(\pv_1,\pv_2,\pv_3)={}&i [p_{1,\mu}Z(\pv_1)-p_{2,\mu}Z(\pv_2)  - (p_{1,\mu}-p_{2,\mu})	\pv_1\cdot\pv_2 D_1[Z](\pv_1,\pv_2) ] T^a_{i_1 i_2}, \label{eq:vert_21_cond}
    \\
    \Gamma_{i_1 i_2 i_3,\mu}^{(3,1)a}(\pv_1,\pv_2,\pv_3,\pv_4)={}&0,
    \\ 
    \Gamma_{i_1 i_2, \mu  \nu}^{(2,2)ab}(\pv_1,\pv_2,\pv_3,\pv_4)={}&  [(T^aT^b)_{i_1 i_2} X_2(\pv_1+\pv_3)+(T^aT^b)_{i_2 i_1} X_2(\pv_1+\pv_4) ] ( \pv_3\cdot \pv_4 \delta_{\mu \nu}-p_{3,\nu} p_{4,\mu} ) \nnl{}&{}
    - \delta_{\mu\nu} [ (T^aT^b)_{i_1 i_2}Z(\pv_1+\pv_3)+(T^bT^a)_{i_1 i_2}Z(\pv_1+\pv_4)-\{T^a,T^b\}_{i_1 i_2}D_1[Z](\pv_1,\pv_2) ] \nnl
    &-(2p_{1,\mu}+p_{3,\mu})p_{1,\nu}(T^aT^b)_{i_1 i_2}D_1[Z](\pv_1,\pv_1+\pv_3)+\text{perm(}(1,2)\text{ or }(3,4)\text{)}\nnl
    &-\dfrac{1}{2}\pv_1\cdot \pv_2 (2p_{1,\mu}+p_{3,\mu})(2p_{2,\nu}+p_{4,\nu})(T^aT^b)_{i_1 i_2}D_2[Z](\pv_1,\pv_2,\pv_1+\pv_3)+\text{perm}(3,4). \label{eq:vert_22_cond}
\end{align}
In Eqs. \eqref{eq:vert_21_cond} and \eqref{eq:vert_22_cond} we have introduced ``discrete derivatives'' defined by
\begin{align}
D_1 [f](\pv_1,\pv_2)& =\dfrac{f(\pv_1)-f(\pv_2)}{\pv_1^2-\pv_2^2},\\
D_2 [f](\pv_1,\pv_2,\pv_3)& =2\dfrac{f(\pv_1)(\pv_2^2-\pv_3^2)+f(\pv_2)(\pv_3^2-\pv_1^2)+f(\pv_3)(\pv_1^2-\pv_2^2)}{(\pv_2^2-\pv_1^2)(\pv_3^2-\pv_2^2)(\pv_1^2-\pv_3^2)},
\end{align}
and verifying
\begin{align}
D_1 [f](\pv,\pv)={}&\partial_{\pv^2}f(\pv),\\
D_2 [f](\pv,\pv,\pv')={}&2\dfrac{f(\pv)-f(\pv')-(\pv^2-\pv'^2)\partial_{\pv^2}f(\pv)}{(\pv^2-\pv'^2)^2},\\
D_2 [f](\pv,\pv,\pv)={}&\partial_{\pv^2}^2f(\pv).
\end{align}

\subsection{Flow equations}
\label{app_subsec_floweq}

\subsubsection{Two-point correlation function}

Restoring the $\rho$ dependence of the functions, one has
\begin{align}
\partial_k W (\rho)  ={}& \tilde{\partial}_k \frac{1}{2} \int_\qv \Big\{ \GL(\qv,\rho) \big[\qv^2 Y(\qv^2)+2 \rho  W''(\rho)+3 W'(\rho)\big]+(N-1) W'(\rho) \GT(\qv,\rho) \Big\} ,
\end{align}
where $\tilde{\partial}_k=(\dk R_k)\partial_{R_k}$ acts only on the $k$ dependence of the cutoff function $R_k$.
Through the remainder of this Appendix all $\rho$-dependent quantities (the propagators, the potential and its derivatives and $\rho$ itself) are evaluated at the running minimum of the potential $\rho_{0,k}$.
\begin{align}
\partial_k Z (\p) ={}& \tilde{\partial}_k \frac{1}{4}  \int_\qv \Big\{ \GT(\qv) \big[2 \rho  \GL(\qv) (\qv^2 Y(\qv)+2 W')^2 -\rho  \GL(\xiv
   ^2) (2 W'+\xiv ^2 Y(\xiv))^2-2 \qv^2 Y(\qv)+2 \xiv ^2 Y(\xiv)\big]\nnl
&{}-\rho 
   \GL(\qv) \GT(\xiv) (\qv^2 Y(\qv)+2 W')^2\Big\},\\
\partial_k Y (\p) ={}&  \tilde{\partial}_k \frac{1}{4 \rho \pv^2} \int_\qv \Big\{ \rho  \Big[4 W' \big(3 \GL(\qv) \big[4 \rho  W'' \GL(\qv)-\GL(\xiv) (\pv^2
   Y(\pv)+4 \rho  W'')\big]-(N-1) \pv^2 Y(\pv) \GT(\qv) \GT(\xiv)\big)
    \nnl
   &{}
   +4
   (W')^2 \Big[\GT(\qv) \big(\GL(\xiv)+(N-1) \big[\GT(\qv)-\GT(\xiv)\big]\big)		
   \nnl
   &{}
   +\GL(\qv) [-9 \GL(\xiv)-2 \GT(\qv)+\GT(\xiv)]+9
   \GL(\qv){}^2\Big]
   	\nnl
   &{}
   +\GL(\qv) \big[16 \rho ^2 (W'')^2 \GL(\qv)-\GL(\xiv)
   (\pv^2 Y(\pv)+4 \rho  W'')^2\big]-(N-1) \pv^4 Y(\pv)^2 \GT(\qv) \GT(\xiv)\Big]
   	\nnl 
   & -2 \qv^2 Y(\qv) \Big(\GL(\qv) \Big[\rho  \big(\GL(\xiv) \big[\pv^2
   Y(\pv)+4 \rho  W''+\xiv ^2 Y(\xiv)\big]   		
   \nnl 
   &
   {}-2 W' \big[-3 \GL(\xiv)+6 \GL(\qv)-2
   \GT(\qv)+\GT(\xiv)\big]-8 \rho  W'' \GL(\qv)\big)+1\Big]-\GT(\qv)\Big)
   \nnl 
   &
   {}+2 \xiv ^2
   Y(\xiv) \Big[\GL(\qv) \big(1-\rho  \GL(\xiv) \big[\pv^2 Y(\pv)+4 \rho  W''+6
   W'\big]\big)+\GT(\qv) \big[2 \rho  W' \GL(\xiv)-1\big] \Big]
   \nnl 
   &
   {}
   +\xiv ^4 \rho  Y(\xiv)^2
   \GL(\xiv) \big[\GT(\qv)-\GL(\qv)\big]+\rho  \qv^4 Y(\qv)^2 \GL(\qv)
   \big[-\GL(\xiv)+4 \GL(\qv)-2 \GT(\qv)+\GT(\xiv)\big]\Big\}.
\end{align}
where for the sake of concision we have defined $\xiv=\pv+\qv$ and for all vectors $\qv^4=|\qv|^4$.

\subsubsection{Scalar susceptibility}

\begin{align}
\partial_k f (\p) ={}&  \tilde{\partial}_k \frac{1}{4} f(\pv)  \int_\qv \big\{- \GL(\xiv) \GL(\qv) \big[\pv^2 Y(\pv)+\qv^2
   Y(\qv)+4 \rho  W''+6 W'+\xiv ^2 Y(\xiv)\big]\nnl
   &{}-(N-1) \GT(\qv) \GT(\xiv) (\pv^2 Y(\pv)+2 W') \big\},\\
\partial_k \gamma (\p) ={}&   -\tilde\partial_k\frac{1}{2} f(\pv)^2 \int_\qv  \big\{\GL(\xiv) \GL(\qv)+(N-1) \GT(\qv) \GT(\xiv) \big\}.
\end{align}

\subsubsection{Conductivity}

\begin{align}
\partial_k X_1 (\p) ={}& \frac{1}{(D-1) \pv^4} \tilde\partial_k \int_\qv \Big\{ \GT(\qv) \Big(\pv^2 \Big[2 \qv^2 \big( (D-1) D_1[Z](\qv,\qv)+2 \big[2 D_1[Z](\qv,\xiv)+D_2[R](\qv,\qv,\xiv)\big]\big)
    \nnl 
   &
    {}+2 (D-1) D_1[R](\qv,\qv)+(D-1) \pv^2 X_2(\xiv)+(D-1) Z(\xiv)\Big]
    \nnl 
   &
    {}-(\xiv ^2-\pv^2-\qv^2)^2 \big[2 D_1[Z](\qv,\xiv)+D_2[R](\qv,\qv,\xiv)\big]+4 \qv^2 \big[\pv^2 \qv^2- (\xiv ^2-\pv^2-\qv^2)^2/4\big]
   D_2[Z](\qv,\qv,\xiv)\Big)
    \nnl 
   &
    {}+\GT(\xiv) \Big(\GT(\qv) \big[(\xiv^2-\pv^2-\qv^2)^2/4-\pv^2 \qv^2\big] \Big[2 \big(\big[ (\xiv ^2-\pv^2-\qv^2)/2+\qv^2\big] D_1[Z](\qv,\xiv)+D_1[R](\qv,\xiv)\big)
    \nnl 
   &
    {}+Z(\qv)+Z(\xiv)\Big]^2+(D-1) \pv^2 \big[\pv^2
   X_2(\qv)+Z(\qv)\big]\Big) \Big\},\\
\partial_k X_2 (\p) ={}& \frac{-1}{4 (D-1) \pv^4 \rho } \tilde\partial_k \int_\qv \Big\{ -(D-1) \Big(D_1[Z]^2(\qv,\xiv) \GL(\xiv) \GT(\qv)
   \qv^4-D_2[Z](\qv,\qv,\xiv) \GL(\qv) \qv^2
   \nnl &
   {}+D_2[Z](\qv,\qv,\xiv) \GT(\qv) \qv^2+(-\pv^2-\qv^2+\xiv ^2) D_1[Z]^2(\qv,\xiv) \GL(\xiv) \GT(\qv) \qv^2
   \nnl &
   {}+2
   Z(\xiv) D_1[Z](\qv,\xiv) \GL(\xiv) \GT(\qv) \qv^2+2 D_1[R](\qv,\xiv) D_1[Z](\qv,\xiv) \GL(\xiv) \GT(\qv) \qv^2
   \nnl &
   {}+X_2(\qv) \GL(\xiv)+X_2(\xiv) \big[\GL(\qv)-\GT(\qv)\big]+Z(\xiv)^2 \GL(\xiv) \GT(\qv)+D_1[R]^2(\qv,\xiv) \GL(\xiv)
   \GT(\qv)
   \nnl &
   {}+[(-\pv^2-\qv^2+\xiv ^2)^2/4] D_1[Z]^2(\qv,\xiv) \GL(\xiv) \GT(\qv)+2 Z(\xiv) D_1[R](\qv,\xiv) \GL(\xiv)
   \GT(\qv)
   \nnl &
   {}+(-\pv^2-\qv^2+\xiv ^2) Z(\xiv) D_1[Z](\qv,\xiv) \GL(\xiv)
   \GT(\qv)+D_2[R](\qv,\qv,\xiv)
   \big[\GT(\qv)-\GL(\qv)\big]
   \nnl &
   {}+(-\pv^2-\qv^2+\xiv ^2) D_1[R](\qv,\xiv) D_1[Z](\qv,\xiv) \GL(\xiv) \GT(\qv)+\Big[\big(Z(\xiv)+D_1[R](\qv,\xiv)
   \nnl &
   {}+\big[\qv^2+ (-\pv^2-\qv^2+\xiv ^2)/2\big] D_1[Z](\qv,\xiv)\big)^2
   \big[\GL(\qv)-2 \GT(\qv)\big]-X_2(\qv)\Big] \GT(\xiv)\Big)
   \pv^4
   \nnl &
   {}+\Big[- (D-1) D_1[Z]^2(\qv,\xiv) \big(\GL(\xiv)
   \GT(\qv)+\big[\GL(\qv)-2 \GT(\qv)\big] \GT(\xiv)\big) (-\pv^2-\qv^2+\xiv^2)^3/2
   \nnl &
   {}-D_1[Z](\qv,\xiv) \big(2 (2 D-3) D_1[Z](\qv,\xiv) \qv^2+(D-1)
   \big[Z(\qv)+3 Z(\xiv)+4 D_1[R](\qv,\xiv)\big]\big) 
   \nnl &
   {}\times\big(\GL(\xiv)
   \GT(\qv)+\big[\GL(\qv)-2 \GT(\qv)\big] \GT(\xiv)\big) (-\pv^2-\qv^2+\xiv^2)^2/2
   \nnl & 
   {}-\Big(2 (D-3) D_1[Z]^2(\qv,\xiv) \big(\GL(\xiv)
   \GT(\qv)+\big[\GL(\qv)-2 \GT(\qv)\big] \GT(\xiv)\big) \qv^4
   \nnl &
   {}+D_1[Z](\qv,\xiv) \Big[\GT(\qv) \big(\big[(D-3) Z(\qv)+(3 D-5) Z(\xiv)+4 (D-2) D_1[R](\qv,\xiv)\big] \big[\GL(\xiv)-2 \GT(\xiv)\big] \qv^2      
   \nnl &
   {}   +2 (D-1)\big)+\GL(\qv) \big(\big[(D-3)
   Z(\qv)+(3 D-5) Z(\xiv)+4 (D-2) D_1[R](\qv,\xiv)\big] \GT(\xiv) \qv^2-2
   D+2\big)\Big]
   \nnl &
   {}-(D-1) \Big[2 D_2[Z](\qv,\qv,\xiv) \big[\GL(\qv)-\GT(\qv)\big] \qv^2+2
   D_2[R](\qv,\qv,\xiv) \big[\GL(\qv)-\GT(\qv)\big]
   \nnl & 
   {}-\big[Z(\xiv)+D_1[R](\qv,\xiv)\big] \big[Z(\qv)+Z(\xiv)+2 D_1[R](\qv,\xiv)\big] \big(\GL(\xiv) \GT(\qv)
      \nnl &
   {}+\big[\GL(\qv)-2 \GT(\qv)\big] \GT(\xiv)\big)\Big]\Big) (-\pv^2-\qv^2+\xiv ^2)    
   \nnl &
   {}+\qv^2 \Big(4 D_1[Z]^2(\qv,\xiv)
   \big(\GL(\xiv) \GT(\qv)+\big[\GL(\qv)-2 \GT(\qv)\big] \GT(\xiv)\big)
   \qv^4 
   \nnl &
   {}-4 D_2[Z](\qv,\qv,\xiv) \GL(\qv) \qv^2+4 D_2[Z](\qv,\qv,\xiv) \GT(\qv)
   \qv^2+Z(\qv)^2 \GL(\xiv) \GT(\qv)
   \nnl &
   {}+Z(\xiv)^2 \GL(\xiv) \GT(\qv)+4
   D_1[R]^2(\qv,\xiv) \GL(\xiv) \GT(\qv) 
   \nnl &
   {}+2 Z(\qv) Z(\xiv)\GL(\xiv) \GT(\qv)+4 Z(\qv) D_1[R](\qv,\xiv) \GL(\xiv)
   \GT(\qv) 
   \nnl &
   {}+4 Z(\xiv) D_1[R](\qv,\xiv) \GL(\xiv) \GT(\qv)+4
   D_2[R](\qv,\qv,\xiv) \big[\GT(\qv)-\GL(\qv)\big]   
   \nnl &
   {}+\big[Z(\qv)+Z(\xiv)+2 D_1[R](\qv,\xiv)\big]^2 \big[\GL(\qv)-2 \GT(\qv)\big] \GT(\xiv)  
   \nnl &
   {}+4 D_1[Z](\qv,\xiv) \Big[ \big[Z(\qv)+Z(\xiv)+2 D_1[R](\qv,\xiv)\big] \big[\GL(\qv)-2 \GT(\qv)\big] \GT(\xiv) \qv^2-2\GL(\qv)
   \nnl &
   {}+\big(\big[Z(\qv)+Z(\xiv)+2 D_1[R](\qv,\xiv)\big] \GL(\xiv) \qv^2+2\big) \GT(\qv)\Big]\Big)\Big] \pv^2   
   \nnl &
   {}-(D/4) (-\pv^2-\qv^2+\xiv ^2)^2 \Big(-4
   D_2[Z](\qv,\qv,\xiv) \GL(\qv) \qv^2+4 D_2[Z](\qv,\qv,\xiv) \GT(\qv)
   \qv^2
   \nnl &
   {}+Z(\qv)^2 \GL(\xiv) \GT(\qv)+Z(\xiv)^2 \GL(\xiv) \GT(\qv)+4
   D_1[R]^2(\qv,\xiv) \GL(\xiv) \GT(\qv)
   \nnl &
   {}+2 Z(\qv) Z(\xiv)
   \GL(\xiv) \GT(\qv)+4 Z(\qv) D_1[R](\qv,\xiv) \GL(\xiv)
   \GT(\qv)
   \nnl &
   {}+4 Z(\xiv) D_1[R](\qv,\xiv) \GL(\xiv) \GT(\qv)+4
   D_2[R](\qv,\qv,\xiv) \big[\GT(\qv)-\GL(\qv)\big]
   \nnl &
   {}+\big[Z(\qv)+Z(\xiv)+2 D_1[R](\qv,\xiv)\big]^2 \big[\GL(\qv)-2 \GT(\qv)\big] \GT(\xiv)
   \nnl &
   {}+ \big[2\qv^2+ (-\pv^2-\qv^2+\xiv ^2)\big]^2 D_1[Z]^2(\qv,\xiv)
   \big(\GL(\xiv) \GT(\qv)+\big[\GL(\qv)-2 \GT(\qv)\big] \GT(\xiv)\big)
   \nnl & 
   {}+4 D_1[Z](\qv,\xiv) \Big[-2 \GL(\qv)+\big(\big[\qv^2+(-\pv^2-\qv^2+\xiv^2)/2\big] 
   \nnl &
   {}\times\big[Z(\qv)+Z(\xiv)+2 D_1[R](\qv,\xiv)\big] \GL(\xiv)+2\big) \GT(\qv)
   \nnl &
   {}+\big[\qv^2+ (-\pv^2-\qv^2+\xiv ^2)/2\big] \big[Z(\qv)+Z(\xiv)+2 D_1[R](\qv,\xiv)\big] \big[\GL(\qv)-2 \GT(\qv)\big] \GT(\xiv)\Big]\Big)\Big\}  ,
\end{align}
where $D=d+1=3$.

\subsection{Dimensionless variables}
\label{app_adim}

The flow equations are solved using the dimensionless variables,
\begin{equation}
\begin{gathered}
\tilde \p=k^{-1} \p, \quad \trho = v_D^{-1} Z_k k^{2-D} \rho , \\
\tilde h = Z_k^{-1} k^{-2} h , \quad \tilde A_\mu = k^{-1} A_\mu, \quad \tilde F_{\mu\nu} = k^{-2} F_{\mu\nu} 
\end{gathered}
\end{equation}
and functions
\begin{equation}
\begin{gathered}
\tilde{W}_k(\trho) =Z_k^{-1}k^{-2}W_k(\rho),\\
\tilde{Z}_k(\tilde\p) =Z_k^{-1}Z_k(\p),\quad
\tilde{Y}_k(\tilde\p) =v_D Z_k^{-2} k^{D-2} Y_k(\p),\\
\tilde f_k(\tilde\p) = f_k(\p) , \quad \tilde \gamma_k (\tilde\p) = v_D^{-1} Z_k^2 k^{4-D} \gamma_k(\p) , \\ 
\tilde{X}_{1,k}(\tilde\p) =v_D^{-1}k^{4-D} X_{1,k}(\p), \quad
\tilde{X}_{2,k}(\tilde\p) =Z_k^{-1} k^{2} X_{2,k}(\p) ,
\end{gathered}
\end{equation} 
where $D=d+1=3$, $v_D=1/2^{D+1} \pi^{D/2}\Gamma(D/2)$ a numerical factor introduced for convenience and $Z_k\equiv Z_k(\p=0)$.

\end{widetext}


\begin{thebibliography}{63}
\bibitem [{not()}]{not1}%
  \BibitemOpen
  \href@noop {} {}\bibinfo {note} {In the quantum model, the stiffness is
  defined by $\chi_{\rm T}(\p,i\wn=0)=2\rho_0/\rho_s\p^2$ for $\p\to
  0$.}\BibitemShut {Stop}%
\end{thebibliography}

%

\end{document}